\documentclass{article}

\usepackage{arxiv}

\usepackage[utf8]{inputenc} 
\usepackage[T1]{fontenc}    
\usepackage{hyperref}       
\usepackage{url}            
\usepackage{booktabs}       
\usepackage{amsfonts}       
\usepackage{nicefrac}       
\usepackage{microtype}      
\usepackage{lipsum}
\usepackage{graphicx}
\graphicspath{ {./images/} }

\usepackage{glossaries}
\setacronymstyle{long-short}
\newacronym{ML}{ML}{machine learning}
\newacronym{XAI}{XAI}{explainable artificial intelligence}
\newacronym[longplural={counterfactual explanations}]{CFE}{CFE}{counterfactual explanation}
\newacronym{AMT}{AMT}{Amazon Mechanical Turk}
\newacronym{GDPR}{GDPR}{General Data Protection Regulation}
\newacronym{MSE}{MSE}{mean squared error}
\newacronym{BMBF}{BMBF}{Federal Ministry of Education and Research Germany}
\newacronym{DLR}{DLR}{German Aerospace Centre}

\usepackage[backend=biber,
            sorting=nyt,
            maxcitenames=2,
            maxbibnames=8,
            backref=true,
            natbib=true]{biblatex}
\usepackage{xspace}
\newcommand*{\eg}{e.g.\@\xspace}
\newcommand*{\ie}{i.e.\@\xspace}

\usepackage{subcaption}
\usepackage{makecell}
\usepackage{booktabs}
\usepackage{multirow}
\usepackage{tabularx}

\usepackage[toc,page,titletoc]{appendix}

\title{Let's Go to the Alien Zoo: Introducing an Experimental Framework to Study Usability of Counterfactual Explanations for Machine Learning}

\author{
Ulrike Kuhl \\
  \textit{CITEC} \\
  Bielefeld University\\
  Bielefeld, Germany \\
  \texttt{ukuhl@techfak.uni-bielefeld.de} \\
   \And
 Andr\'e Artelt \\
  \textit{CITEC} \\
  Bielefeld University\\
  Bielefeld, Germany \\
  \texttt{aartelt@techfak.uni-bielefeld.de} \\
  \And
 Barbara Hammer \\
  \textit{CITEC} \\
  Bielefeld University\\
  Bielefeld, Germany \\
  \texttt{bhammer@techfak.uni-bielefeld.de} \\
}

\begin{document}
\maketitle
\begin{abstract}
To foster usefulness and accountability of \gls{ML}, it is essential to explain a model's decisions in addition to evaluating its performance.
Accordingly, the field of \gls{XAI} has resurfaced as a topic of active research, offering approaches to address the ``how'' and ``why'' of automated decision-making.
Within this domain, \glspl{CFE} have gained considerable traction as a psychologically grounded approach to generate post-hoc explanations.
To do so, \glspl{CFE} highlight what changes to a model's input would have changed its prediction in a particular way.
However, despite the introduction of numerous \gls{CFE} approaches, their usability has yet to be thoroughly validated at the human level.
Thus, to advance the field of \gls{XAI}, we introduce the Alien Zoo, an engaging, web-based and game-inspired experimental framework. 
The Alien Zoo provides the means to evaluate usability of \glspl{CFE} for gaining new knowledge from an automated system, targeting novice users in a domain-general context.
As a proof of concept, we demonstrate the practical efficacy and feasibility of this approach in a user study. 
Our results suggest that users benefit from receiving \glspl{CFE} compared to no explanation, both in terms of objective performance in the proposed iterative learning task, and subjective usability.
With this work, we aim to equip research groups and practitioners with the means to easily run controlled and well-powered user studies to complement their otherwise often more technology-oriented work.
Thus, in the interest of reproducible research, we provide the entire code, together with the underlying models and user data: \url{ https://github.com/ukuhl/IntroAlienZoo}
\end{abstract}

\glsresetall
\section{Introduction}\label{sec:introduction}

\insert\footins{
  \normalfont\footnotesize
  \interlinepenalty\interfootnotelinepenalty
  \splittopskip\footnotesep \splitmaxdepth \dp\strutbox
  \floatingpenalty10000 \hsize\columnwidth
  \includegraphics[scale=0.5]{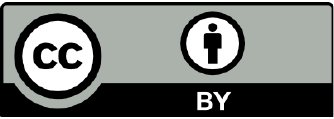} This work is licensed under a Creative Commons \href{https://creativecommons.org/licenses/by/4.0/}{``Attribution 4.0 International''} license.}

In a step towards accountable and transparent \gls{ML}, the European Union mandates safeguards against automated decision-making with the \gls{GDPR} 2016/679 \cite{GDPR16}. 
Specifically, Recital 71 of the \gls{GDPR} states that a person subjected to automated decision-making may obtain an explanation of the given decision. 
While it is debated whether this establishes a legally binding `right to explanation' \cite{wachter2017right}, it has certainly sparked lively discussion about scope, realization, and feasibility of explanations for \gls{ML} in the domain of \gls{XAI}. Consequently, there has been an upswing of technical \gls{XAI} approaches on how to make \gls{ML} explainable in recent years~\cite{arrieta2020explainable,chou2022counterfactuals}.

Alongside novel explainability approaches, authors have proposed evaluation criteria and guidelines to systematically characterize and assess \gls{XAI} approaches \cite{arrieta2020explainable,davis2020measure,doshi-velez_towards_2017,sokol2020explainability}.
For instance, \citeauthor{sokol2020explainability} suggest a taxonomy defining criteria an explanatory method has to satisfy to be considered usable, summarized in an \textit{``Explainability Fact Sheet''}.

This theoretical groundwork sparked generation of several practical validation frameworks, focusing on function level validation of explanation approaches~\citep{arras2022clevr,artelt_evaluating_2021,de_oliveira_framework_2021,pawelczyk2021carla,sattarzadeh2021svea}.
For instance, these frameworks evaluate explanations in terms of their accuracy and fidelity~\citep{arras2022clevr,pawelczyk2021carla,sattarzadeh2021svea,white_measurable_2020}, or robustness~\citep{artelt_evaluating_2021}.

Importantly, however, these considerations pass over the role of the user as eventual target - a curious limitation, given that user studies are considered the gold standard in \gls{XAI} \citep{doshi-velez_towards_2017,sokol2020explainability}.
While \gls{XAI} taxonomies repeatedly emphasize the need for human-level validation of explanation approaches, only few authors concern themselves with user-based evaluations, often with limitations concerning statistical power and reproducibility~\citep{keane_if_2021}.

The repeated emergence of counter-intuitive findings from sparse user evaluations acts as a stark reminder why accounting for the human factor is vital when evaluating \gls{XAI} approaches.
For instance,~\citeauthor{poursabzi2021manipulating} show that participants can more easily simulate predictions of clear models with few features, however, this does not lead users to adjust their behavior more closely in line with the model's predictions. 
Moreover, instead of the expected advantage of clear models in terms of their interpretability, users are actually less able to detect when the model had made a mistake~\citep{poursabzi2021manipulating}.
Similarly, employing the Alien Zoo framework introduced in this manuscript, we demonstrate that introducing a theoretically motivated plausibility constraint on generated explanations may be less useful for users in certain settings~\citep{kuhl2022keep}.

What inhibits systematic and controlled comparisons of XAI approaches from a usability perspective?
While user evaluations are essential to evaluate the efficacy of explanation modes, designing an effective user study is no easy feat. 
A well-designed study needs to closely consider the respective explainees, and the reason for explaining \citep{adadi2018peeking, sokol2020explainability}, while simultaneously taking into account confounding factors and available resources \citep{doshi-velez_towards_2017}.
Further, it is challenging to ensure comparability of conditions human participants face, while systematically varying \gls{XAI} approaches, underlying \gls{ML} models, or data distributions.

Thus, lack of openly accessible and engaging user study designs that enable direct comparisons between different explainability implementations, models, and data sets motivates the current work.
To advance the field of \gls{XAI}, we introduce the Alien Zoo, an engaging, web-based and game-inspired experimental framework. 
The Alien Zoo provides means to evaluate the usability of a specific and very prominent variant of post-hoc, model agnostic explanations for \gls{ML}, namely \glspl{CFE}~\citep{artelt_computation_2019}, targeting novice users in a domain-general context.

The aim of this contribution is to equip research groups and practitioners to with an easily adaptable design, adjustable for various purposes and research questions.
With this paradigm, we account for a series of challenges that are sometimes overlooked in previous \gls{XAI} user studies, overcoming prominent shortcomings in the literature (see Section \ref{subsec:DesignPrinc}). 
Thus, we aspire to narrow the gulf between the increasing interest in generating human-friendly explanations for automated decision-making, and the limitations given current user-based evaluations.

As a proof of concept, we demonstrate the efficacy and feasibility of the Alien Zoo approach in a user study, showing a beneficial impact of providing \glspl{CFE} on user performance in the proposed iterative learning task.
Providing the entire code, together with the underlying data and scripts used for statistical evaluation\footnote{Available at \url{ https://github.com/ukuhl/IntroAlienZoo}}, our hope is that this framework will be utilized by other research groups and practitioners.

The remainder of this paper is as follows: 
We will first provide a primer on \glspl{CFE} for \gls{ML}, and briefly review previous usability assessments and respective lessons learned (Section \ref{sec:CFEs}).
Subsequently, we will detail our the conceptualization of the proposed Alien Zoo framework, including guiding design principles, constructs and measurements, and implementation specifics (Section \ref{sec:AlienZooPar}).
Section \ref{sec:EmpStudy} describes the accompanying proof of concept usability study, demonstrating the efficacy and feasibility of the Alien Zoo approach to evaluate \glspl{CFE}.
We close this paper with an in-depth discussion of insights drawn from this study, including limitations and avenues for future work in Section \ref{sec:Disc}.

\section{CFEs for ML}\label{sec:CFEs}

The advent of novel \gls{XAI} approaches also triggered a shift toward a user-centered focus on explainability~\citep{miller_explanation_2019}.
In the wake of this change in focus, \glspl{CFE} received special attention as a supposedly useful, human-friendly solution ~\citep{keane_if_2021,miller_explanation_2019}.
\glspl{CFE} for \gls{ML} correspond to \textit{what-if} scenarios, highlighting necessary changes in a model's input that trigger a desired change in the model's output (\ie, ``if you earned US\$ 200 more per month, your loan would be approved'').

What makes \glspl{CFE} so attractive as an \gls{XAI} approach?
Foremost, their contrastive nature, emphasizing why a specific outcome occurred instead of another, strongly resembles human cognitive reasoning~\citep{hilton_knowledge-based_1986,lipton_contrastive_1990,lombrozo_explanation_2012,miller_explanation_2019}.
Humans naturally reflect on past events by generating possible alternatives, thus routinely engaging in counterfactual thinking~\citep{roese_counterfactual_1997}.
Specifically, when humans generate counterfactuals, they take the relevant facts of events as input, and mentally change their mental representation of these facts in order to produce a counterfactual scenario, while maintaining the factual representation in parallel~\citep{byrne_counterfactual_2016}.

Empirical evidence shows that humans engage in counterfactual thinking automatically~\citep{goldinger_blaming_2003}.
In a series of experiments,~\citeauthor{sanna_antecedents_1996} demonstrate that participants produce counterfactual thoughts spontaneously in various settings (\eg, when re-telling stories, evaluating their own performance on an exam, or describing their performance in a laboratory anagram task)~\citep{sanna_antecedents_1996}. 
Importantly, the number of spontaneously uttered counterfactuals increases when participants face negative outcomes or results were unexpected ~\citep{sanna_antecedents_1996}.

Based on these and similar insights,~\citeauthor{epstude_functional_2008} emphasize the beneficial role of this mode of thinking to regulate one's behavior in order to improve future performance~\citep{epstude_functional_2008}. 
In the updated version of their \textit{Functional Theory of Counterfactural Thinking}, they posit that disparities between the current state and an ideal goal state triggers spontaneous counterfactual thought~\citep{roese_functional_2017}.
In the same vein,~\citeauthor{markman_reflection_2003} argue that comparative thought modes like generating counterfactuals may help to prepare for the future by guiding the formation of intentions and thus changing prospective behavior~\citep{markman_reflection_2003}.
Taken together, the contrastive and spontaneously way humans generate counterfactuals that presumably guide future behavior leads many authors in \gls{XAI} to treat explanations formulated as counterfactuals as intuitively useful, and readily human-usable~\citep{artelt_convex_2020,dandl2020multi,guidotti_local_2018,stepin_paving_2019}.

However, these encouraging insights from psychology seem to have created the erroneous impression that technical \glspl{CFE} approaches providing explainability for \gls{ML} models may take the quality of the suggested explanation modes at face value~\citep{doshi-velez_towards_2017, offert_i_2017}.
According to a recent review, only one in three counterfactual \gls{XAI} papers include user-based evaluations, often with limited statistical validity and little opportunity to reproduce the presented results~\citep{keane_if_2021}.

\subsection{Insights from Previous CFE Usability Studies}

While still a minority, studies that do examine \gls{CFE} approaches from a user-perspective give cause for cautious confidence in their usability.
~\citeauthor{van_der_waa_evaluating_2021} demonstrate that \glspl{CFE}, compared to example based and no-explanation control variants, enable users interacting with a hypothetical decision support system for diabetes patients to correctly identify features relevant for a system’s prediction~\citep{van_der_waa_evaluating_2021}
Additionally, their data suggest that \glspl{CFE} have a positive effect on perceived system comprehensibility compared to no explanation.
Focusing on the issue of fairness,~\citeauthor{dodge2019explaining} find that counterfactual explanations most prominently expose biased classifiers, compared to alternative approaches like showing feature relevance or merely describing the distribution of underlying data~\citep{dodge2019explaining}.
Finally, a comparison of different explanation approaches reveals that participants judge counterfactual style explanations to be subjectively more intuitive and useful than, \eg, visualizing feature importance scores \citep{le2020grace}.

This positive evidence is not unanimous, however. Users tasked to learn how an automatic system behaves indeed show some understanding of the types of rules governing said system after receiving counterfactual-style explanations, as compared to receiving no-explanation
~\citep{lim2009and}. Yet, only participants that are presented with explicit feedback stating why the system behaved in a certain way perform consistently better across several metrics, including perceived understanding. 
\citeauthor{lage2019human} demonstrate that users show consistently higher response times when asked to answer counterfactual style questions, indicating increased cognitive load for this type of task, a factor that may actually hinder usability~\citep{lage2019human}.

On top of these inconsistent results, previous user evaluations often suffer from a series of limitations.
While it is intuitively clear that one explanation mode fitting all scenarios is unlikely to exist~\citep{sokol2020one}, 
some authors neglect to clearly formulate the given purpose for explaining and target group in their study~\citep{le2020grace}. Without an explicit classification of the experimental context, research runs the risk to reach all-too-general conclusions like declaring one mode of explaining universally superior to another.

Many user evaluations suffer from methodological limitations like low participant numbers~\citep{akula2020cocox,lim2009and}.
For reasons of simplicity, some approaches provide participants with explanations that follow a certain \gls{XAI} approach, but were actually designed by the researchers themselves \citep{lage2019human, narayanan2018humans, van_der_waa_evaluating_2021}.
Such a \textit{Wizard of Oz} approach, with a human behind the scenes plays the role of an automatic system~\citep{dahlback1993wizard}, allows perfect control over materials encountered by participants. However, it fails to account for potential variability in the results of \gls{ML} algorithms. For instance, it is perfectly conceivable that an approach may produce unhelpful explanations in certain settings. 
If this may happen to users `in the wild', we posit that assessment of these approaches in the lab needs to account for such contingencies as well. 

Another prominent limitation affects evaluations that merely focus on assessing perceived usability. Using questionnaires and surveys is a prominent approach to ask participants how well they like or understand a certain explainability method. However, it is unclear whether such subjective evaluations translate into tangible behavioral effects \citep{hoffman2018metrics}. In fact, a recent study fails to show a correlation between perceived system understandability and any objective measure of performance \citep{van_der_waa_evaluating_2021}.

Further, experimental designs in XAI are often limited in terms of engaging participants, especially in studies assessing the efficacy of explanations for understanding how a system works.  
For instance, participants typically study pre-selected examples of a system's input and output values, together with a corresponding explanation~\citep{le2020grace, lim2009and, van_der_waa_evaluating_2021}. 
However, greater focus on user action may be advantageous. 
Evidence from educational science suggests that learner's level of commitment relates to final learning outcomes, with interactive activities granting deeper understanding \citep{chi_icap_2014}.

Last, many designs reported are difficult to exactly reproduce as experimental code, \gls{ML} models, and underlying data are not openly available. This lack of shared resources severely hampers replication studies and adaptation of frameworks according to novel research purposes near to impossible.

Thus, while more and more empirical evaluations assessing the usability of \glspl{CFE} for \gls{ML} take the scene, experimental shortcomings become apparent in a number of ways. 
Moreover, the diverse range of different set-ups and paradigms, testing different \gls{CFE} methods in different scenarios and with diverse groups of users, impeding firm conclusions in terms of strengths and weaknesses of methods, as studies plainly lack comparability.

\section{The Alien Zoo Framework}\label{sec:AlienZooPar}

\subsection{Guiding Design Principles}\label{subsec:DesignPrinc}

As more and more user studies in the domain of \gls{XAI} emerge, more and more recommendations and guidelines concerning design principles that need to be taken into account enter the picture~\citep{davis2020measure,mohseni2021multidisciplinary,van_der_waa_evaluating_2021}.
For instance, based on their experiences setting up an \gls{XAI} usability study, \citeauthor{van_der_waa_evaluating_2021} formulate recommendations as a reference for future \gls{XAI} research~\citep{van_der_waa_evaluating_2021}.
We closely followed these guidelines when constructing the Alien Zoo framework. 

\subsubsection{Use case and experimental context}

\begin{figure}[t]
   \centering
   \includegraphics[width=\columnwidth]{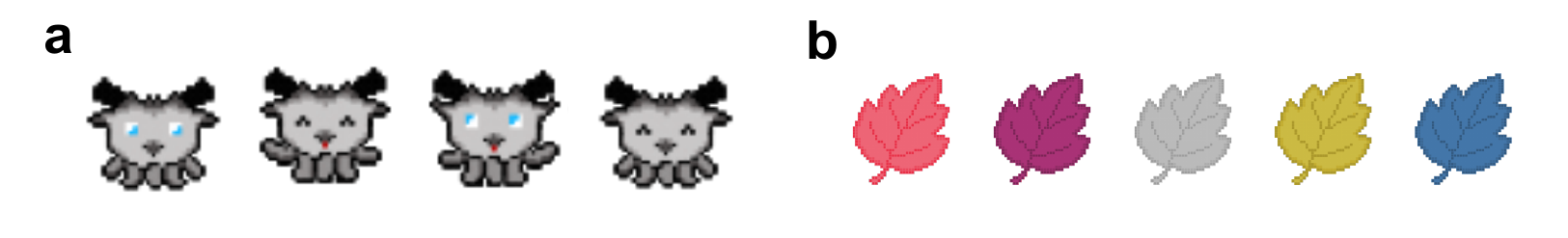}
   \caption{Integral components of the Alien Zoo framework: (\textbf{a}) An exemplary group of shubs, the small alien species inhabiting the zoo. (\textbf{b}) Plants available to the participants for feeding.}
   \label{fig:ShubsPlants}
 \end{figure}

The effectiveness of an explanation depends decisively on the reason reason for explaining, and the intended target audience~\citep{adadi2018peeking,arrieta2020explainable,mohseni2021multidisciplinary}.
Both aspects determine the choice of an appropriate use case, and thus the experimental context.

On the one hand, a user's background knowledge crucially impacts their judgement whether a piece of information is relevant: 
Users who already possess a lot of applicable domain knowledge may find more sophisticated explanations more useful than users that are novices~\citep{doshi-velez_towards_2017}.
Even more critically, prior domain knowledge and user beliefs may impact how and even if users meaningfully engage with provided explanations~\citep{lim2009and}.
Moreover, explainees equipped with AI expertise perceive and evaluate provided explanations differently than users that lack this kind of knowledge~\citep{ehsan2021explainable}.

On the other hand, the explanation's purpose profoundly affects requirements a given \gls{XAI} approach ought to meet.
Users tasked to compare different models likely have other explanation needs than those who want to gain new knowledge from a predictive model or the data used to build it~\citep{adadi2018peeking}.

Consequently, generalizability of conclusions beyond a given use case, context, and target group is limited and needs to be treated with upmost caution~\citep{doshi-velez_towards_2017,sokol2020explainability}.

Prominent \gls{XAI} taxonomies take these aspects into consideration.  
For instance, the previously mentioned \textit{``Explainability Fact Sheet''} includes characteristics of context (i.e., \textit{Explanation audience} and \textit{Function of the Explanation}) as part of the operational dimension of the \gls{XAI} evaluation~\citep{sokol2020explainability}.  
\citet{doshi-velez_towards_2017} provide a structured classification of evaluation approaches for interpretability, differentiating between application-grounded, human-grounded, and functionally-grounded approaches.

In the Alien Zoo framework, we focus on an abstract experimental context:
Participants imagine themselves as zookeepers in a zoo for aliens, so-called shubs (Figure \ref{fig:ShubsPlants}a). 
Participants' main task is to find how to best feed the shubs under their care. 
They may choose from different plants to feed the aliens (Figure \ref{fig:ShubsPlants}b), but it is not clear what plants (or which plant combination) makes up a nutritious diet, causing their pack to thrive.
Feeding decisions have immediate consequences, leading to reductions or increases of the pack size.
In regular intervals, participants receive \glspl{CFE} together with their past choices, highlighting an alternative selection that would have led to a better result. Thus, the current use case is that of assisting novice users without any prior experience to gain new knowledge from a predictive model about the data used to build it.

With the Alien Zoo framework, we provide a highly interactive, game-like scenario that triggers participant engagement over iterative rounds of user action and feedback.
Evidence from educational science motivates this choice, demonstrating that learner's level of commitment affects learning outcome, and that interactive activities foster deeper understanding~\citep{chi_icap_2014}.

In the following we adhere to the advice by~\citeauthor{van_der_waa_evaluating_2021} and first provide a structured account of the choice for a use case following well-defined taxonomies. Subsequently, we discuss the effectiveness of the use case domain for the intended purpose of the evaluation, and review our choice for running the proof of concept investigation as a web-based study~\citep{van_der_waa_evaluating_2021}.

With the given use case, we assess performance of real users in an abstract task setting. 
Thus, Alien Zoo user studies correspond to human grounded evaluations, with participants engaged in a variant of ``counterfactual simulation''~\citep{doshi-velez_towards_2017}. As described by~\citeauthor{adadi2018peeking}, our setting falls into the ``explaining to discover'' category for explainability, evaluating whether providing \glspl{CFE} to novice users enhances their ability to extract yet unknown relationships within an unfamiliar dataset~\citep{adadi2018peeking}.

A particular advantage concerns the abstract nature of our task, ruling out any potential confounding effects of prior user knowledge: 
it is safe to say that any user is a novice when it comes to feeding aliens, eliminating the possibility of misconceptions or prior beliefs. 

Last, we diverge from the final recommendation by~\citeauthor{van_der_waa_evaluating_2021} regarding the experimental context: instead of using a controlled, lab-based setting, our framework is a web-based design. While we understand concerns regarding validity of data acquired this way~\citep{van_der_waa_evaluating_2021}, we demonstrate the feasibility of using the given online approach to obtain meaningful data if appropriate quality measures are in place (see Section \ref{subsec:dat-qual}).

\subsubsection{Constructs and their relations}

\begin{figure}[t]
   \centering
   \includegraphics[width=\columnwidth]{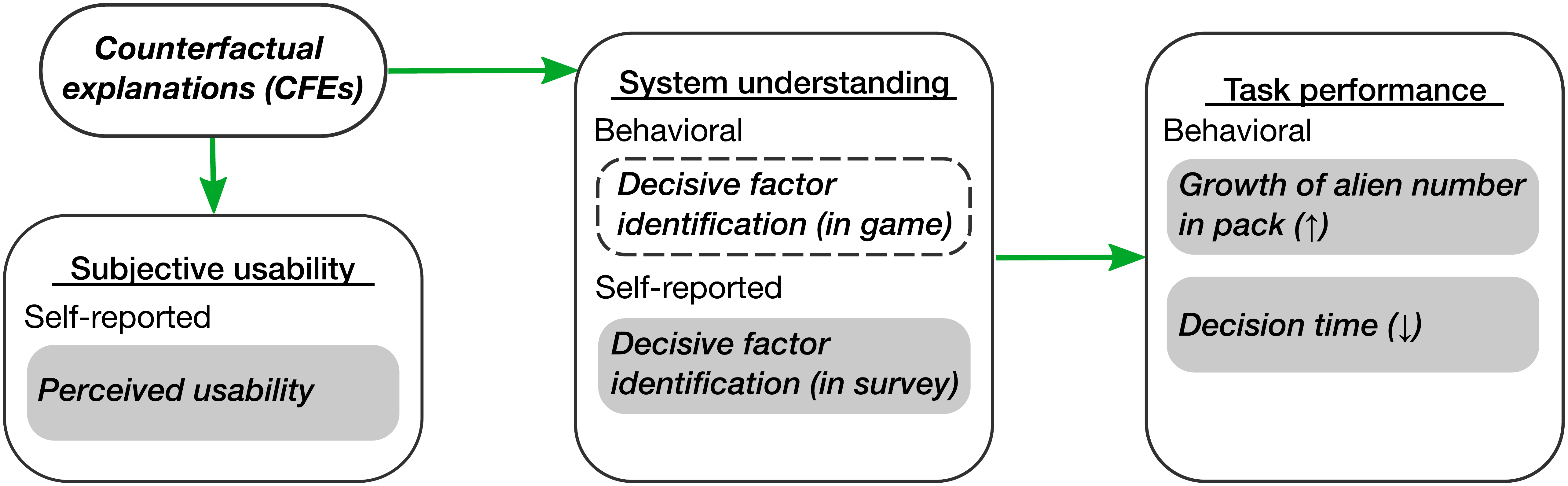}
   \caption{Causal diagram describing expected effects of counterfactual explanations on the constructs subjective usability, system understanding, and task performance investigated in the Alien Zoo framework.
   Green arrows depict expected positive effects.
   Opaque gray boxes show the measures for the respective construct, divided into behavioral and self-reported measurements. 
   Arrows behind measures depict the expected direction of positive effects.
   The dashed box shows a behavioral factor reflecting improved system understanding indirectly assessed via its mediating role on task performance.}
   \label{fig:CausalDiagramm}
\end{figure}

To underlay any \gls{XAI} user evaluation with a solid basis for scientific theory,~\citeauthor{van_der_waa_evaluating_2021} advocate clear definitions of utilized constructs and their interrelations~\cite{van_der_waa_evaluating_2021}.

Alien Zoo user evaluations focus on three constructs: subjective usability, system understanding, and task performance.
Figure \ref{fig:CausalDiagramm} depicts a causal diagram showing the expected relations between these constructs.
Specifically, we posit that providing \glspl{CFE} positively impacts a user's system understanding, as well as their subjective usability. 
Consequently, increased system understanding will enable users to better perform the task at hand.

The given proof of concept study described in Section \ref{sec:EmpStudy} compares user performance when receiving \glspl{CFE} with a no explanation a control. 
When provided with \glspl{CFE}, we expect participants to gain a better understanding of decisive features, and the best combination thereof, in the underlying data. 
Consequently, we anticipate increased system understanding to improve task performance.
Given how humans engage in counterfactual thinking automatically on a day-to-day basis~\citep{goldinger_blaming_2003,roese_counterfactual_1997,sanna_antecedents_1996}, we expect that explanations formulated as counterfactuals also have a positive impact on subjective understanding. 

Finally, it is crucial to consider subjective usability as a construct separate from system understanding.
A participant's action does not necessarily correspond to their perceived system understanding, strongly suggesting that user behavior and self-report do not measure the same construct~\citep{van_der_waa_evaluating_2021}.

\subsubsection{Measurements}
The Alien Zoo framework enables assessment of constructs subjective usability, system understanding, and task performance through different behavioral and self-report measures (Figure \ref{fig:CausalDiagramm}).
Note that we rely on both objective behavioral variables and subjective self-reports.
Given evidence of a disparity between perceived system understandability and objective measures of performance~\citep{van_der_waa_evaluating_2021}, we believe it crucial to address both aspects in order to provide a holistic assessment of usability of \gls{CFE} for \gls{ML}.

First, we hypothesize that providing \glspl{CFE} to show participants alternative feeding choices leading to better outcomes triggers better understanding of the system. 
Specifically, we expect participants to recall and apply the information provided by \glspl{CFE} to improve their feeding choice. 
Ultimately, this translates to an increase in the participant’s capacity to correctly identify the decisive factors in the data used to train the shub growth model, both in the study game and survey phase. 
While this capacity is not directly measured during the game, we acquire corresponding self-reports via the post-game survey.
The first two survey items assess whether users can identify those plants that contribute to successful completion of a task, and those that do not matter to make the pack grow. Thus, we determine to what extent people have an explicit understanding of the data structure.

Second, we expect that system understanding has a positive effect on task performance.
Measures assessing task performance include the number of aliens in the pack over the duration of the game (henceforth referred to as pack size).
This value indirectly quantifies the extent of user's understanding of relevant and irrelevant features in the underlying data set, as a solid understanding leads to better feeding choices. 
Further, we expect time needed to reach a feeding decision over trials to be indicative of how well participants can work with the Alien Zoo (henceforth referred to as decision time). As we assume participants to become more automatic in making their plant choice, we expect this practice effect to be reflected as decreased decision time~\citep{logan_shapes_1992}. 

Last, self-reports acquired via the post-game survey assess different aspects of how participants judge the subjective usability of explanations provided (for a full list of all survey items, see Supplementary Material \ref{app:suppSurvey}).
For instance, users indicate whether they understood the explanations, in how far they find them useful, to which degree they can make use of them, and in how far they imagine the presented \glspl{CFE} to be helpful for other users, too. These items assess user's subjective usability.

\subsection{System Implementation}

The implementation of the Alien Zoo realizes a strict separation of the front end creating the game interface participants interact with, and the back end providing the required \gls{ML} functionality.
The web interface employs the JavaScript-based Phaser 3, an HTML5 game framework.\footnote{\url{https://phaser.io/}}
The back end of the system is Python3-based,
with the sklearn package~\citep{pedregosa_scikit-learn_2011} supporting \gls{ML} processes.
An underlying \gls{ML} model trained on synthetic plant data to predict the alien pack's growth rate determines the behavior of the game.
This model receives input from the user end to update the current number of shubs. 
To ensure flexibility in terms of potential models, we employ the CEML toolbox~\cite{artelt_ceml_2019} to compute \glspl{CFE}.\footnote{\url{https://github.com/andreArtelt/ceml}}
CEML is a Python toolbox for generating \glspl{CFE}, supporting many common machine learning frameworks to ensure availability of a wide range of potential \gls{ML} algorithms.
Thus, the Alien Zoo provides a highly flexible infrastructure to efficiently investigate different intelligibility factors of automatically generated \glspl{CFE}.

The web-based nature of the infrastructure allows for prompt data collection of a large number of participants. 
In an associated study investigating the effects of plausibility constraints on generated \glspl{CFE}, we acquired data from over 100 participants within four days via \gls{AMT}~\citep{kuhl2022keep}. Data acquisition from 90 participants in the current proof of concept study (Section \ref{sec:EmpStudy}) took five days, including the initial quality assessment.

Unlike previous designs that provide users with hand-crafted explanation examples~\citep[so-called \textit{Wizard of Oz} designs, see][]{lage2019human,narayanan2018humans,sokol2020one,van_der_waa_evaluating_2021}, the Alien Zoo equips participants with feedback from real \gls{XAI} methods based on reproducible \gls{ML} models. 
We agree that \textit{Wizard of Oz} designs are the preferable option in terms of control over what participants experience and consistency of presented explanations~\citep[]{browne2019wizard,jentzsch2019conversational}.
However, we believe that only a test of explanations that genuinely come from an ML model can show whether such explanations are useful, re-creating how users would interact with them `in the wild'.

In the interest of reproducibility, we fully share data and code of the Alien Zoo framework, encouraging research groups and practitioners alike to adapt and utilize the implementation according to their own research needs.

\section{Empirical Proof of Concept Study}\label{sec:EmpStudy}

In the following, we empirically investigate the efficacy and feasibility of the Alien Zoo framework.
To this effect, to employ it to run a user study examining the impact of providing \glspl{CFE} on user performance as compared to no explanations in the proposed Alien Zoo iterative learning task.
The study consists of two experiments that primarily vary in terms of the complexity of the underlying data used for model building.
Specifically, growth rate in Experiment 1 depends on the best combination of three plants, while this is reduced to the best combination of two plants in Experiment 2 (see Section \ref{subsubsec:models}).
The critically low learning rate of control participants in Experiment 1 motivated our decision to run the second experiment.
Thus, we investigated whether users that do not receive explanations generally fail to learn in the Alien Zoo setting, even given a simpler configuration.

\subsection{Hypotheses}\label{sec:hypotheses}

The guiding question of the empirical proof of concept study is whether users benefit from receiving \glspl{CFE} when tasked to identify relationships within an unknown data set when interacting with the Alien Zoo framework. 

We evaluate this question using an interactive iterative learning task, in which users repeatedly select input values for an \gls{ML} model.
Throughout the experiment, users receive feedback at regular intervals.
Either we show them an overview of their choices alone (control condition), or we show them this overview alongside \glspl{CFE}, highlighting how changes in their past choices may have led to better results (\glspl{CFE} condition).
Via this approach, the interaction between repeated actions by users and corrective feedback allows us to assess system understanding objectively through task performance over a series of decisions.

We hypothesize that providing \glspl{CFE} compared to no explanations indeed helps users in the task at hand.
Specifically, we assume that exposure to alternative feeding choices that would lead to better results enables users to build a more accurate mental model of the underlying data distribution.

We recruited novice users and designed the task around an abstract scenario in order to gain insight into the usability of \glspl{CFE}.
By using this approach, we can protect against possible differences in domain knowledge or misconceptions about the task setting that might impact task performance~\citep{van_der_waa_evaluating_2021}.

Consequently, we address the following three hypotheses.

\paragraph{Hypothesis 1}
We expect users that receive \glspl{CFE} on top of a summary of their past choices to outperform users without explanations in discovering unknown relationships in data, both in terms of objective and subjective measures.
Specifically, we anticipate that participants in the \glspl{CFE} condition a) produce larger pack sizes, thus showing greater learning success, b) become faster in making their choice as a sign of more automatic processing, and c) are able to explicitly identify relevant and irrelevant input features.

\paragraph{Hypothesis 2}
In terms of subjective understanding, we predict a marked group difference.
We expect that users that receive \glspl{CFE} will subjectively find their feedback more helpful and usable. 
Furthermore, we posit that those users will also judge \gls{CFE} feedback to be more helpful for other users.

\paragraph{Hypothesis 3}
Both feedback variants (overview of past choices and overview of past choices + \glspl{CFE}) are relatively straight-forward. 
Thus, when evaluating users' understanding of the feedback themselves, and their evaluation of timing and efficacy of how feedback is presented, we do not expect to see group differences.
Further, it will be interesting to see if users differ in terms of needing support to understand the provided feedback.
In the \gls{CFE} condition, it is conceivable users may wish for additional help for interpreting this added information.

\subsection{Methods}

\subsubsection{Participants}

We conducted the study in early March 2022 on \gls{AMT}. We restricted access to the study to users that (a) belong to the high performing workers on the platform, and have been granted the Mechanical Turk Masters Qualification, (b) have a work approval rate of at least 99\%, and (c) did not participate before in any Alien Zoo tasks we ever ran on \gls{AMT}.

For each experiment, we recruited 45 participants, randomly assigned to either \textit{CFE} or \textit{control} (\ie, no explanation) group. All participants gave informed electronic consent by providing click wrap agreement prior to participation.
Participants received payment after first data quality assessment. 

Contributions from participants whose data showed insufficient quality (see Section \ref{subsec:dat-qual}) were rejected.
Affected users received US\$ 1 base compensation for participation, paid via the bonus system. 
This concerned 6/45 (Experiment 1) and 2/45 (Experiment 2) participants, respectively.
All remaining participants received a base pay of US\$ 3 for participation. 
The five best performing users in each experiment received an additional bonus of US\$ 1. 
We included information about the prospect of a bonus in the experimental instructions, to motivate users to comply with the task~\citep{bansal_updates_2019}.
The Ethics Committee of Bielefeld University, Germany, approved this study.

\subsubsection{Experimental Procedure}\label{subsec:experimental-procedure}
The experiment consists of a game and a survey phase.
Accepting the task on \gls{AMT} redirects participants to a web server hosting the study.

Users are first notified of the purpose, procedure and expected duration of the study, their right to withdraw, confidentiality and contact details of the primary investigator.
If a user does not wish to participate, they may close the window.
Otherwise, users confirm their agreement via button press.
As soon as they indicate agreement, participants get secretly allotted to one of the experimental conditions (either \textit{control} or \textit{\gls{CFE}}) via random assignment.

A subsequent page provides detailed information about the Alien Zoo game.
Specifically, it illustrates images of the aliens to be fed, and the variety of plants they may use for feeding.
Written instructions state that a pack size can be increased or decreased by choosing healthy or unhealthy combinations of leaves per plant. 
The maximal number of leaves per plant is limited to six, and users may freely select any combination of plants they find preferable. 
Subsequent instructions direct the user to maximize the number of aliens, so-called shubs, in order to qualify as a top player to receive an additional monetary bonus.
Further, written information establishes that participants will receive a summary of their past choices after two rounds of feeding.
Users in the \glspl{CFE} condition also learn that they will be provided with feedback on what choice would have led to a better result on these occasions.

Clicking a ``Start'' button at the end of the page indicates that the user is ready to start the game phase.
This button appears with a delay of 20s in an effort to prevent participants from skipping the instructions.

\paragraph{Game Phase}

\begin{figure*}[ht]
   \centering
   \includegraphics[width=\textwidth]{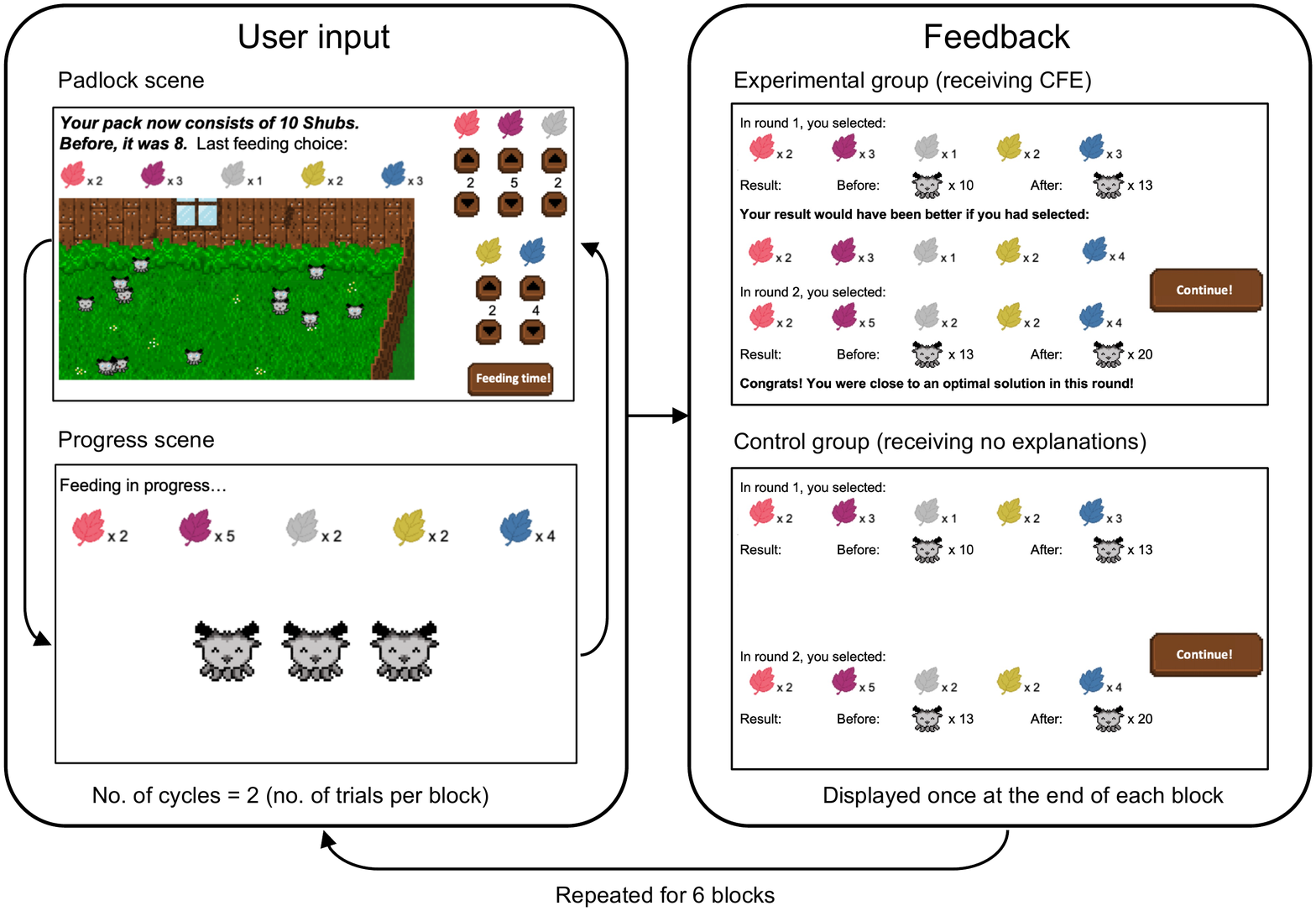}
   \caption{General flow of scenes displayed during the game phase. Note that this pattern was disrupted after trials 3 and 7 for an additional attention scene, asking participants to indicate the current number of shubs in their pack. A more detailed overview of scenes within a block can be found in Supplementary Material \ref{app:suppSceneFlow}.}
   \label{fig:GamePhaseFlow}
 \end{figure*}

Figure \ref{fig:GamePhaseFlow} visualizes the general flow of scenes displayed during the game phase. This phase begins with a padlock scene, where participants make their first feeding selection (left side in Figure \ref{fig:GamePhaseFlow}, and Supplementary Material \ref{app:suppSceneFlow}).
All available plant types alongside upward and downward arrow buttons appear on the right side of this scene. 
The same leaf icon in different colors represents the different plants (Figure \ref{fig:ShubsPlants}b). 
While each participant encounters the same 5 plant colors, their order is randomized for each participant in order to avoid confounding effects.
During the first trial, the top of the page notes that clicking on the upward and downward arrows increases and decreases the number of leaves of a specific plant, respectively.
In each subsequent trial, the top of the page holds a summary of the previous trial's choice, together with the previous and current pack size.
Furthermore, the page shows a padlock displaying the current pack of animated shubs.
Participants receive a pack of 20 shubs to begin.
Participants submit their choice by clicking a ``Feeding time!'' button in the bottom right corner of the screen.

While users watch a short progress scene, the underlying \gls{ML} model predicts the new growth rate based on the user's input.
Our implementation subsequently updates the pack size based on the model's decision, and computes a \gls{CFE}.
Within three seconds, a new padlock appears, visualizing the impact of the current choice in terms of written information and animated shubs. The choice procedure repeats after odd trials.

Users receive feedback after even trials, accessible via a single ``Get feedback!'' button replacing the choice panel on the right-hand side of the screen.
The feedback button directs users to an overview of past two feeding choices, and the impact on pack size.
Users in the \textit{\gls{CFE}} condition are additionally presented with the intermittently computed \glspl{CFE}, illustrating an alternative choice that would have led to a better result for each of the past two trials. If users select a combination of plants that lead to maximal increase in pack size, no counterfactual will be computed. In these cases, users learn that they were close to an optimal solution in that round.

Hitting a ``Continue!'' button appearing after 10s on the right-hand side of the screen, users proceed with the next trial, encountering a new padlock scene. 
We included this delay to ensure that users spend sufficient time with the presented information to be able to draw conclusions for their upcoming feeding decisions.
Each experiment in this paper consists of 12 trials (\ie, 12 feeding decisions). 
Users receive feedback after even trials.

Two additional attention checks assess attentiveness of users during the game phase, implemented after trials 3 and 7.
Said attention checks request participants to type in the current number of shubs in the last feeding round.
Participants receive immediate feedback on the correctness of their answer, alongside a reminder to stay attentive to every aspect of the game at all times. The game then continues with the subsequent progress scene. 

After the user made 12 feeding decisions, the game phase of the study ends.

\paragraph{Survey Phase}
In the survey phase, users answer a series of questions. Survey items first assess user's explicit knowledge of plant relevance for task success (items 1 and 2), and second subjective judgements of usability and quality of feedback provided via an adapted version of the System Causability Scale \cite{holzinger_measuring_2020}.

A final set of three self-report measures assesses potential confounding factors. 
They address whether users understand the feedback provided, whether they feel they need support for understanding it, and how they evaluate the timing and efficacy of feedback.
The last two items of the survey phase collect demographic information on participant's gender and age.

On the final page of the study, users are thanked for their participation and receive a unique code to provide on the \gls{AMT} platform to prove that they completed the study and qualify for payment.
To ensure anonymity, we encrypt payment codes and delete them as soon as users received payment.

Finally, participants may choose to follow a link providing full debriefing information.

\subsubsection{Data Quality Criteria}\label{subsec:dat-qual}

Due to the nature of web-based studies, some users may attempt to game the system, claiming payment without providing adequate answers.
Thus, a priori defined criteria ensure sufficient data quality.

Users qualify as speeders based on their decision time in the padlock scene, if they spent less than two seconds to make their plant selection in at least four trials.
Users qualify as inattentive participants if they fail to give the correct number of shubs in both attention trials (game phase). Likewise, we categorize participants as inattentive users if they fail to select the requested answer when responding to the catch item in the survey phase (see Supplementary Material \ref{app:suppSurvey}). Finally, users qualify as straight-liners if they keep choosing the same plant combination despite not improving in three blocks or more (game phase), or if they answer with only positive or negative valence in the survey phase.

By excluding data of individuals that were flagged for at least one of these reasons from further analysis, we maintain a high level of data quality.

\subsubsection{Statistical Analysis}
We perform all statistical analyses using R-4.1.1~\citep{r_core_team_r_2021}, using experimental condition (\textit{control} and \textit{\gls{CFE}}) as independent variable.
Staying true to our longitudinal design, linear mixed models examine effects of experimental condition over the 12 experimental trials (lme4 v.4\textunderscore 1.1-27.1)~\citep{bates_fitting_2015}.
In the model evaluating differences in terms of user performance, number of shubs generated serves as dependent variable.
In the model evaluating differences in terms of user's reaction time, decision time in each trial serves as dependent variable.
Each model includes the fixed effects of group, trial number and their interaction.
The random-effect structure includes a by-subjects random intercept. 
We decided to follow this approach, as linear mixed models account for correlations of data drawn from the same participant~\citep{detry_analyzing_2016,muth_alternative_2016}.
To compare model fits, we rely on the analysis of variance function of the stats package in base R.
$\eta_{\text{p}}^{2}$ values denote effect sizes (effectsize v.0.5)~\citep{ben-shachar_effectsize_2020}.
We follow up significant main effects or interactions by computing pairwise estimated marginal means, with respective effect sizes reported in terms of Cohen's \textit{d}.
To account for multiple comparisons, all post-hoc analyses reported are Bonferroni corrected.

We evaluate data acquired during the survey phase depending on item type.
The first two items assess user's explicit knowledge of plant relevance, or irrelevance, for task success.

We aim to obtain a unified measure of user knowledge, appreciating correct answers but also penalizing incorrect ones. Therefore, we use the number of matches between user input and ground truth (i.e., number of plants correctly identified as relevant or irrelevant) per participant per item.
Distributions of match data was tested for normality using the Shapiro-Wilk test, followed up by the non-parametric Wilcoxon-Mann-Whitney \textit{U} test in case of non-normality, and the Welch two-sample t-test otherwise for group comparisons. 
We follow the same approach to compare age and gender distributions.
Finally, we gauge group differences of ordinal data from the Likert-style items, using the non-parametric Wilcoxon-Mann-Whitney \textit{U} test.
Effect sizes for all survey data comparisons are given as \textit{r}.

\subsubsection{Models}\label{subsubsec:models}

To predict the growth rate and thus ultimately the new pack size given the user input in each trial, we train a decision tree regression model for each experiment.
Decision trees consecutively split the data along a series of if-then-else rules, thus approximating the underlying data distribution \citep{shalev-shwartz_understanding_2014}.
Decision trees are powerful enough to model our synthetic data set with sufficient accuracy, while allowing for efficient computation of \gls{CFE} \citep{artelt_computation_2019}.\footnote{Note, however, that the Alien Zoo framework itself does not depend on a specific model, and could potentially used with other regression models as well.}
The current implementation uses the Gini splitting rule of CART \citep{breiman_classification_1984}. 
To maintain comparable model outputs for all users within throughout one experiment, we use the same decision tree model once build in the beginning.

\paragraph{Hyperparameter tuning}

To ensure the models reliably present the respective underlying data structure without overfitting, we choose tree depth that yield a high $R^2$ value and minimizes the \gls{MSE} when evaluated on test data.
As a further sanity check, we ensured that inputting the perfect solution into the model reliably yields no CFE (i.e, eliciting the feedback that one is close to an optimal solution). 
Overly complex models are prone to overfit, picking up dependencies in the structure of the randomly chosen features.
\glspl{CFE} generated on the basis of such a model may suggest changes in irrelevant features, thus leading participants on a garden-path.
Thus, for the more complex data set in Experiment 1, we use a maximal tree depth of 7 (model performance on test data: $R^2$ = 0.893, \gls{MSE} = 0.037), while the tree model in Experiment 2 was trained with a maximal tree depth of 5 (model performance on test data: $R^2$ = 0.888, \gls{MSE} = 0.039).

\paragraph{Training Data}

\begin{figure*}[t]
    \centering
    \begin{subfigure}[b]{0.5\textwidth}
        \centering
        \includegraphics[width=\textwidth]{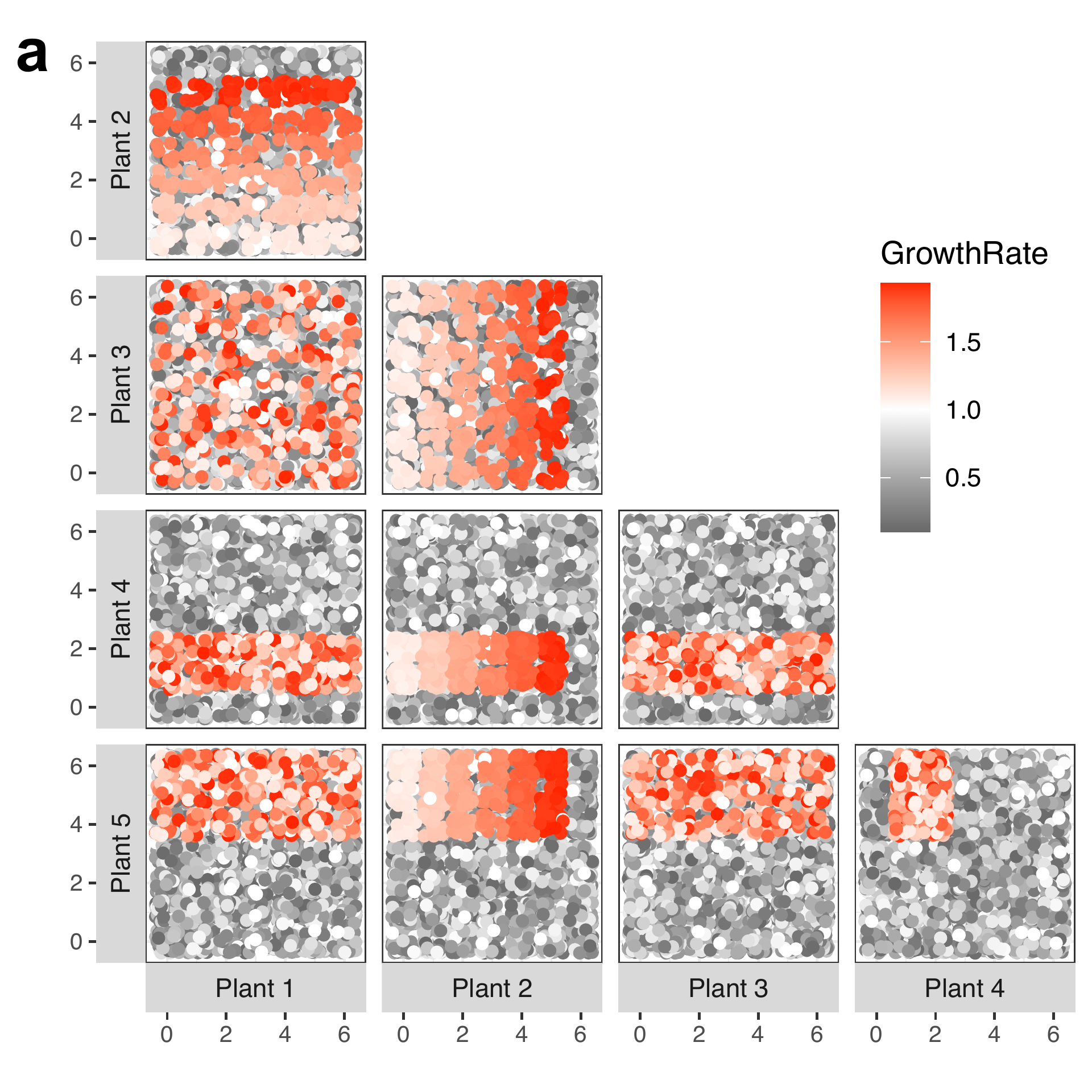}
         \label{fig:DataDistExp1}
    \end{subfigure}%
    \hfill
    \begin{subfigure}[b]{0.5\textwidth}
        \centering
        \includegraphics[width=\textwidth]{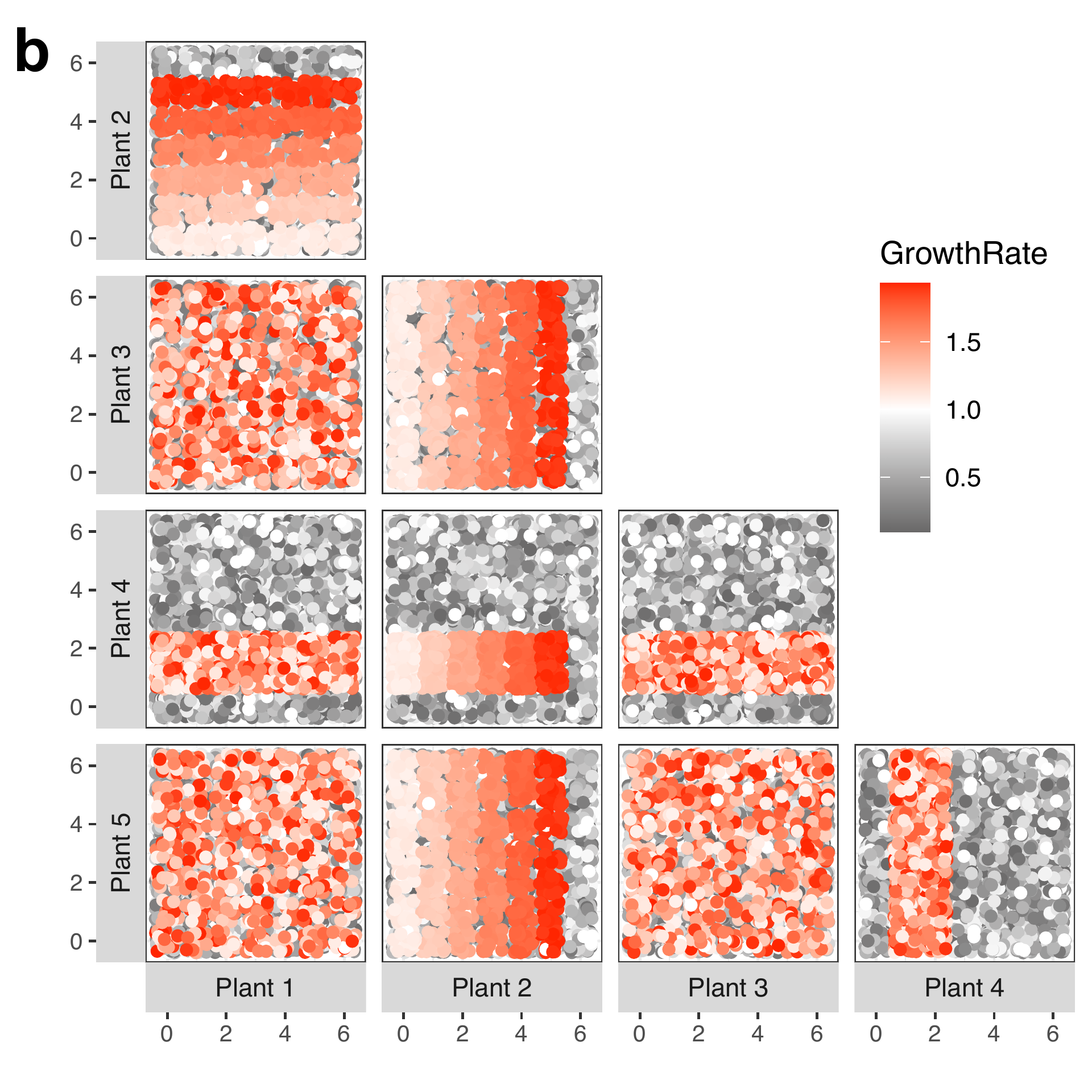}
         \label{fig:DataDistExp2}
     \end{subfigure}
    \caption{Distribution of synthetic data used for model training. Each point in each scatter plot represents the combination of two plant values, colored according to the corresponding growth rate of that point. Gray values indicate growth rate values below 1 (leading to pack size decreases), and red values code values above 1 (leading to pack size increases). (\textbf{a}) Experiment 1: The growth rate scales linearly with plant 2, depending on values of plant 4 and plant 5. (\textbf{b}) Experiment 2: The growth rate scales linearly with plant 2, depending on values of plant 4. For clear rendering, only 0.2\% of all training data are shown, with data points are jittered around their true integer values.}
        \label{fig:DataDists}
\end{figure*}

The underlying data in Experiment 1 were generated according to the following scheme: 
The growth rate scales linearly with values 1 to 5 for plant 2, iff plant 4 has a value of 1 or 2 AND plant 5 is not smaller than 4 (Figure \ref{fig:DataDists}a). 
For Experiment 2, we reduced the dependency to two relevant features, such that growth rate scales linearly with values 1 to 5 for plant 2, iff plant 4 has a value of 1 or 2 (Figure \ref{fig:DataDists}b).
In both experiments, the linear relationship does not hold for value 6 of plant 2, to prevent a simple maximization strategy with respect to this feature.

Growth rate may take a value between 0.1 and 1.9.\footnote{Originally, the prediction was conceived to be used as a factor, enabling exponential growth in perfect cases. This was changed because it meant that individual people might achieve very high pack sizes, in turn disproportionally driving potential effects.} 
In each trial, the respective model predicts the new growth rate based on the current user input.
Subsequently, the new growth rate (range 0.1-1.9) is converted into a corresponding value between -10 and 10 in our implementation, that gets then added to the current number of shubs to update the pack size. Note that our implementation prevents pack size from shrinking below two.

Each synthetic data set contains all possible plant – growth rate combinations 100 times, yielding 1680700 data points. 
For final model training, we balance the data set by first binning the samples based on their label (growth rate), and then applying Synthetic Minority Over-sampling Technique (SMOTE)~\cite{chawla2002smote} using the bins as class labels. The final data set is obtained by removing the binning.

\subsection{Results}
The empirical part of the current paper investigates whether the proposed Alien Zoo framework is suitable to study the effect of providing automatically generated \glspl{CFE} for users tasked to learn about yet unknown relationships in a data set.
We used an abstract setting to circumvent any confounding effects from previous knowledge of the users.

\subsubsection{Experiment 1}

In Experiment 1, we acquired data from 45 participants (Table \ref{tab:Exp1Participants}), tasked to identify relationships within an unknown data set. 
To ensure sufficient task complexity, we opted for a comparatively complex interdependence of three features.

\paragraph{Participant Flow}
From 45 participants recruited via \gls{AMT}, we exclude data from participants who failed both attention trials during the game (n = 2), and straight-lined during the game despite not improving (n = 4). No participant in this cohort qualified as speeder, gave an incorrect response for the catch item in the survey, or straight-lined in the survey. 
Thus, the final analysis includes data from 39 participants (Table \ref{tab:Exp1Participants}).
Note that for one user in the CFE condition, logging of responses for the first two survey items (``Which plants were [not] relevant to increase the number of Shubs in your pack?'') failed.
Thus, we excluded this user in the evaluation of these two items, but included them in all remaining analysis.

On average, the final 39 participants in Experiment 1 needed 17m:42s ($\pm$ 01m:16s SEM) from accepting the task on \glspl{AMT} to inserting their unique payment code.

\begin{table}[tb]
  \caption{Demographic information of participants in Experiment 1.}
  \label{tab:Exp1Participants}
\begin{tabularx}{\textwidth}{llllllllll}
\toprule
    & \multicolumn{4}{l}{Before quality assurance measures (\textit{N} = 45)} & & \multicolumn{4}{l}{After quality assurance measures (\textit{N} = 39)}\\
\cline{2-5} \cline{6-10} 
    & \textit{control} & \textit{CFE} & \textit{U} value$^a$ & \textit{p} value & & \textit{control} & \textit{CFE} & \textit{U} value$^a$ & \textit{p} value \\ 
\hline
\textit{N}   &  22 & 23 & .. & .. & & 19 & 20 & .. & .. \\
Gender$^b$ & 5f/17m & 5f/18m & 255.5 & .950 & & 4f/15m & 5f/15m & 182.5 & .788 \\
Age (\textit{Mdn})$^c$ & 35--44y & 35--44y & 225.5 & .516 & & 35--44y & 35--44y & 143 & .168 \\
\bottomrule
\multicolumn{10}{l}{$^a$ non-parametric Wilcoxon-Mann-Whitney \textit{U} test}\\
\multicolumn{10}{l}{$^b$ f = female, m = male}\\
\multicolumn{10}{l}{\makecell[l]{$^c$ \textit{Mdn} = median age band (options: 18-24y, 25-34y, 35-44y, 45-54y, 55-64y, 65y and over)}}
\end{tabularx}
\end{table}

\begin{figure*}[t]
   \centering
   \includegraphics[width=\textwidth]{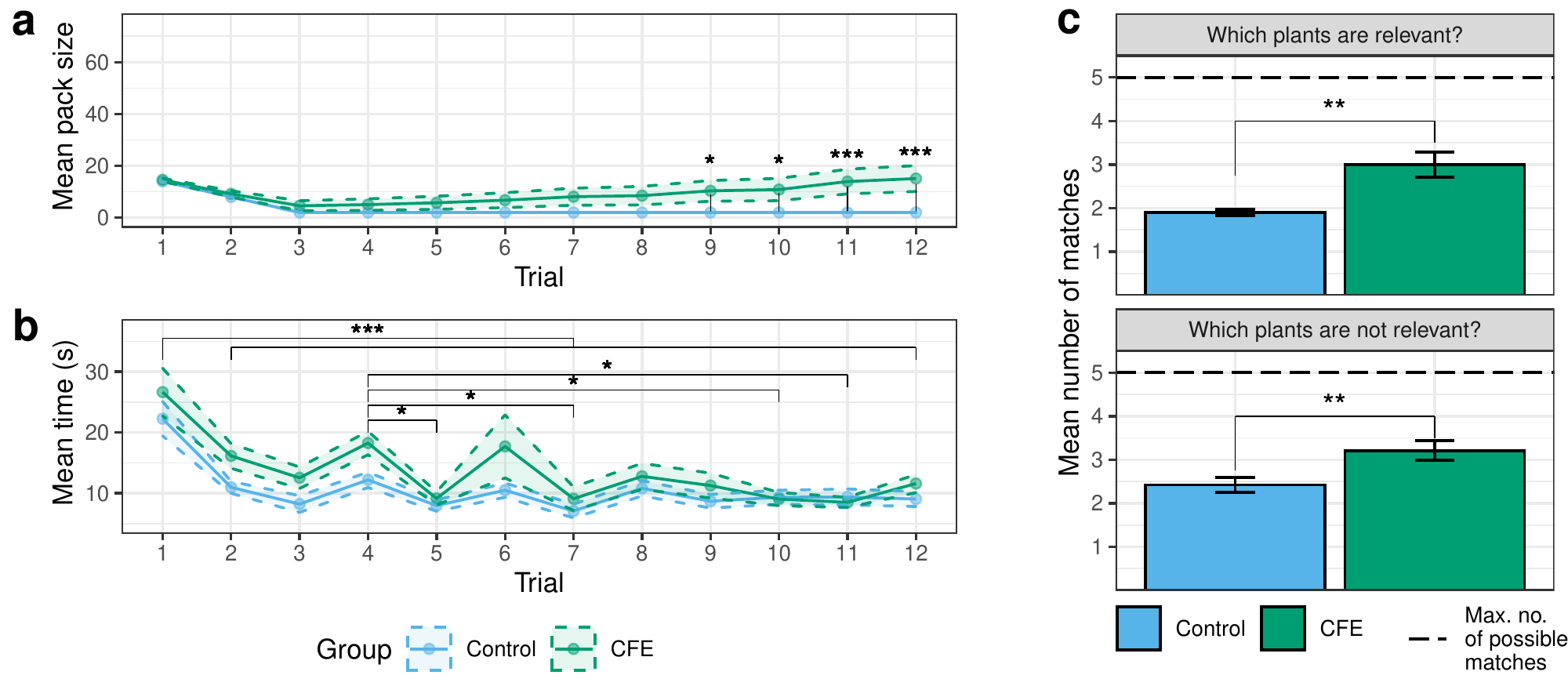}
   \caption{Experiment 1: Development of (\textbf{a}) mean pack size per group by trial, (\textbf{b}) mean decision time per group by trial, and (\textbf{c}) mean number of matches between user judgments and ground truth for survey items assessing relevant plants and irrelevant plants, respectively. Shaded areas in (\textbf{a}) and (\textbf{b}), and error bars in (\textbf{c}) denote the standard error of the mean. Asterisks denote statistical significance (\textit{p} < .05 (*), and \textit{p} < .001 (***), respectively.}
   \label{fig:Exp1Hyp1}
 \end{figure*}

\paragraph{Objective Measures of Usability}
Hypothesis 1 posits that users benefit from receiving \glspl{CFE} compared to no explanations in the Alien Zoo framework.
To address this hypothesis, we compare data from participants in both groups in terms of pack size produced over time, decision time, and matches between ground truth and indicated plants. 
Figure \ref{fig:Exp1Hyp1}a depicts the development of average pack size as well as average decision time per group.
While users receiving CFEs clearly show a positive trajectory, users receiving no explanation did not show any trace of improvement over the course of this experiment.
In fact, no user in the control condition managed to increase their pack size from the minimal attainable number of two by trial 12.
A significant interaction of factors trial number and group (F(11,407) = 6.649, \textit{p} \textless .001, $\eta_{\text{p}}^{2}$ = 0.153) in the corresponding linear mixed effects model confirms this stark discrepancy. 
Follow-up analysis reveal significant differences between groups from trial 9 onward (t(56.7) $\geq$ 2.461, \textit{p} $\leq$ 0.0169, \textit{d} $\geq$ 1.711).
Additionally, there is a significant main effect of trial number (\textit{F}(1,407) = 15.758, \textit{p} \textless .001, $\eta_{\text{p}}^{2}$ = 0.299), but no significant main effect of group (\textit{F}(1,37) = 3.755, \textit{p} = .060, $\eta_{\text{p}}^{2}$ = 0.092).

Participants in either group showed a marked decrease in decision time over the curse of the study, especially after the very first trial (Figure \ref{fig:Exp1Hyp1}b).
A significant main effect of factor \textit{trial number} (\textit{F}(11,407) = 13.025, \textit{p} \textless .001, $\eta_{\text{p}}^{2}$ = 0.260) confirms this observation.
Corresponding post-hoc analyses show significant differences between trial 1 and all other trials (all \textit{t}(407) $\geq$ 5.189, \textit{p} \textless .001, \textit{d} \textgreater 1.175).
Moreover, decision time for trial 4 as the initial trial after the first in-game attention question, stands out. Users require significantly more time to reach a feeding decision in trial 4 compared to trial 5 (\textit{t}(407) = 3.755, \textit{p} = .013, \textit{d} = 0.850), trial 7 (\textit{t}(407) = 4.020, \textit{p} = .005, \textit{d} = 0.911), trial 10 (\textit{t}(407) = 3.397, \textit{p} \textless .049, \textit{d} = 0.769), and trial 11 (\textit{t}(407) = 3.537, \textit{p} \textless .030, \textit{d} = 0.801).
Neither the main effect of factor \textit{group} (\textit{F}(1,37) = 3.976, \textit{p} = .054, $\eta_{\text{p}}^{2}$ = 0.097), nor the interaction between factors \textit{trial number} and \textit{group} (\textit{F}(11,407) = 0.965, \textit{p} = .477, $\eta_{\text{p}}^{2}$ = 0.025) reach significance.

Thus, these results verify our hypothesis that providing \glspl{CFE} in the AlienZoo not just facilitates, but enables learning in the first place, given the poor performance of participants in the control group.

\paragraph{Assessing user's explicit knowledge}
In terms of mean number of matches between user judgments of plant relevance for task success and the ground truth, participants receiving \glspl{CFE} could explicitly identify relevant plants
(\textit{control}: mean number of matches between user input and ground truth = 1.895 $\pm$ 0.072 \textit{SE}; \textit{CFE}: mean number of matches = 3.000 $\pm$ 0.286 \textit{SE}; \textit{U} = 281.5, \textit{p} = .001, \textit{r} = .517) 
as well as irrelevant plants (\textit{control}: mean number of matches between user input and ground truth = 2.421 $\pm$ 0.176 \textit{SE}; \textit{CFE}: mean number of matches = 3.210 $\pm$ 0.224 \textit{SE}; \textit{U} = 264.5, \textit{p} = .009, \textit{r} = .422) more easily than users receiving no explanation. 

\paragraph{Measures of Subjective Usability}

Hypothesis 2 posits that providing CFEs compared to no explanation increases user's subjective understanding.
To assess this notion, we analyze participant judgments on relevant items in the post-game survey.

\begin{figure*}[ht]
   \centering
   \includegraphics[width=\textwidth]{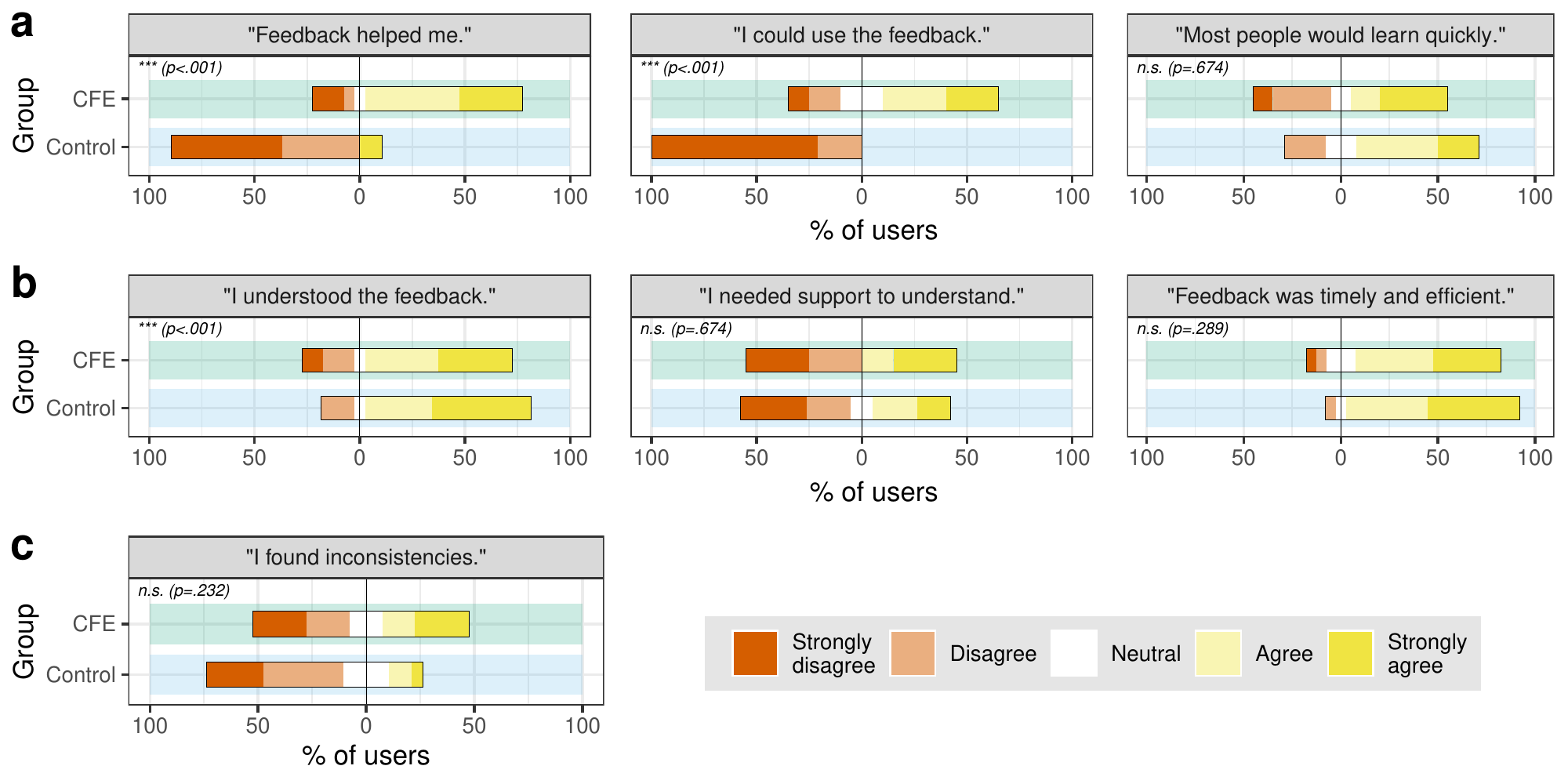}
   \caption{Experiment 1: Overview of user responses in post-game survey (adapted from ~\citep{holzinger_measuring_2020}) per group. (\textbf{a}) depicts user replies in survey items relevant for hypothesis 2, (\textbf{b}) depicts user replies in survey items relevant for hypothesis 3, and (\textbf{c}) depicts replies relevant for our last exploratory analysis. 
   Statistical information including the respective \textit{p}-value is given within each item's box (n.s. = not significant).}
   \label{fig:Exp1Survey}
 \end{figure*}

Visual assessment of user responses suggest large discrepancies between groups in items assessing feedback's helpfulness and usability (Figure \ref{fig:Exp1Survey}a).
This notion is confirmed by the corresponding statistical assessment. 
Groups differ when judging whether presented feedback (\ie, summary of past choices only vs. summary + \glspl{CFE}) was helpful to increase pack size (\textit{control} condition: \textit{M} = 1.789 $\pm$ 0.282 \textit{SE}; \textit{CFE} condition: \textit{M} = 3.700 $\pm$ 1.285 \textit{SE}; \textit{U} = 306.5, \textit{p} \textless .001, \textit{r} = .540).
Similarly, participants receiving \glspl{CFE} on top of a summary of their past choices significantly differed in terms of reported subjective usability (\textit{control} condition: \textit{M} = 1.210 $\pm$ 0.096 \textit{SE}; \textit{CFE} condition: \textit{M} = 3.450 $\pm$ 0.294 \textit{SE}; \textit{U} = 351, \textit{p} \textless .001, \textit{r} = .759).
Strikingly, however, there is no significant difference between groups for estimated usefulness of feedback for others (\textit{control} condition: \textit{M} = 3.632 $\pm$ 0.244 \textit{SE}; \textit{CFE} condition: \textit{M} = 3.350 $\pm$ 0.335 \textit{SE}; \textit{U} = 175, \textit{p} = .674, \textit{r} = .067).

\paragraph{Mode of Presenting Feedback and CFEs}
In conflict with Hypothesis 3, survey responses reflecting user's subjective understanding of feedback show that groups differ in terms of understanding the feedback as such (Figure \ref{fig:Exp1Survey}b). 
While a considerable proportion of both groups responds positively about understanding the feedback, the \textit{control} group leans significantly more to giving positive judgements (\textit{control} condition: \textit{M} = 4.105 $\pm$ 0.252 \textit{SE}; \textit{CFE} condition: \textit{M} = 3.7 $\pm$ 0.309 \textit{SE}; \textit{U} = 312.5, \textit{p} \textless .001, \textit{r} = .567). 
When indicating their need for support for understanding, both groups reply with a comparable, more balanced response pattern (\textit{control} condition: \textit{M} = 2.684 $\pm$ 0.351 \textit{SE}; \textit{CFE} condition: \textit{M} = 2.900 $\pm$ 0.383 \textit{SE}; \textit{U} = 205, \textit{p} = .674 \textit{r} = .067).
User judgements on timing and efficacy of presented feedback is consistently high across groups (\textit{control} condition: \textit{M} = 4.316$\pm$ 0.188 \textit{SE}; \textit{CFE} condition: \textit{M} = 3.950$\pm$0.246 \textit{SE}; \textit{U} = 154.5, \textit{p} = .289 \textit{r} = .170).

\paragraph{Identification of Inconsistencies}
Our explanatory analysis revealed that users in groups did not differ in finding inconsistencies in the feedback provided (\textit{control} condition: \textit{M} = 2.316 $\pm$ 0.265 \textit{SE}; \textit{CFE} condition: \textit{M} = 2.95 $\pm$ 0.352 \textit{SE}; \textit{U} = 232, \textit{p} = .232, \textit{r} = .192).

\subsubsection{Experiment 2}

In Experiment 2, we acquired data from 45 additional participants facing the same task as in Experiment 1 (Table \ref{tab:Exp2Participants}).
The underlying data used for model training was simpler, including the interdependence of two and not three features.

\paragraph{Participant Flow}
From 45 participants recruited via \gls{AMT}, we exclude data from participants who failed both attention trials during the game (n = 1), and straight-lined during the survey (n = 1). No participant in this cohort qualified as a speeder, gave an incorrect response for the catch item in the survey, or straight-lined in the game part of the study. 
Thus, the final analysis includes data from 43 participants (Table \ref{tab:Exp2Participants}).

On average, the these 43 participants in Experiment 2 needed 14m:25s ($\pm$ 01m:07s SEM) from accepting the task on \glspl{AMT} to inserting their unique payment code.

\begin{table}[b]
  \caption{Demographic information of participants in Experiment 2.}
  \label{tab:Exp2Participants}
\begin{tabularx}{\textwidth}{llllllllll}
\toprule
    & \multicolumn{4}{l}{Before quality assurance measures (\textit{N} = 45)} & & \multicolumn{4}{l}{After quality assurance measures (\textit{N} = 43)}\\
\cline{2-5} \cline{6-10}
    & \textit{control} & \textit{CFE} & \textit{U} value$^a$ & \textit{p} value & & \textit{control} & \textit{CFE} & \textit{U} value$^a$ & \textit{p} value\\ 
\hline
\textit{N}   &  21 & 24 & .. & .. & & 21 & 22 & .. & ..\\
Gender$^b$ & 9f/11m/1nb & 11f/13m & 238 & .725 & &  9f/11m/1nb & 9f/13m & 229 & .967\\
Age (\textit{Mdn})$^c$ & 35--44y & 35--44y & 280 & .497 & & 35--44y & 35--44y & 253 & .571\\
\bottomrule
\multicolumn{10}{l}{$^a$ non-parametric Wilcoxon-Mann-Whitney \textit{U} test}\\
\multicolumn{10}{l}{$^b$ f = female, m = male, nb = non-binary}\\
\multicolumn{10}{l}{\makecell[l]{$^c$ \textit{Mdn} = median age band (options: 18-24y, 25-34y, 35-44y, 45-54y, 55-64y, 65y and over)}}
\end{tabularx}
\end{table}

\paragraph{Objective Measures of Usability}
In Experiment 2, we successfully replicate the beneficial effect of providing CFEs compared to no explanations in the Alien Zoo approach already seen in Experiment 1.
This is noteworthy, given the less complex interdependencies within the underlying data set.  

\begin{figure*}[t]
   \centering
   \includegraphics[width=\textwidth]{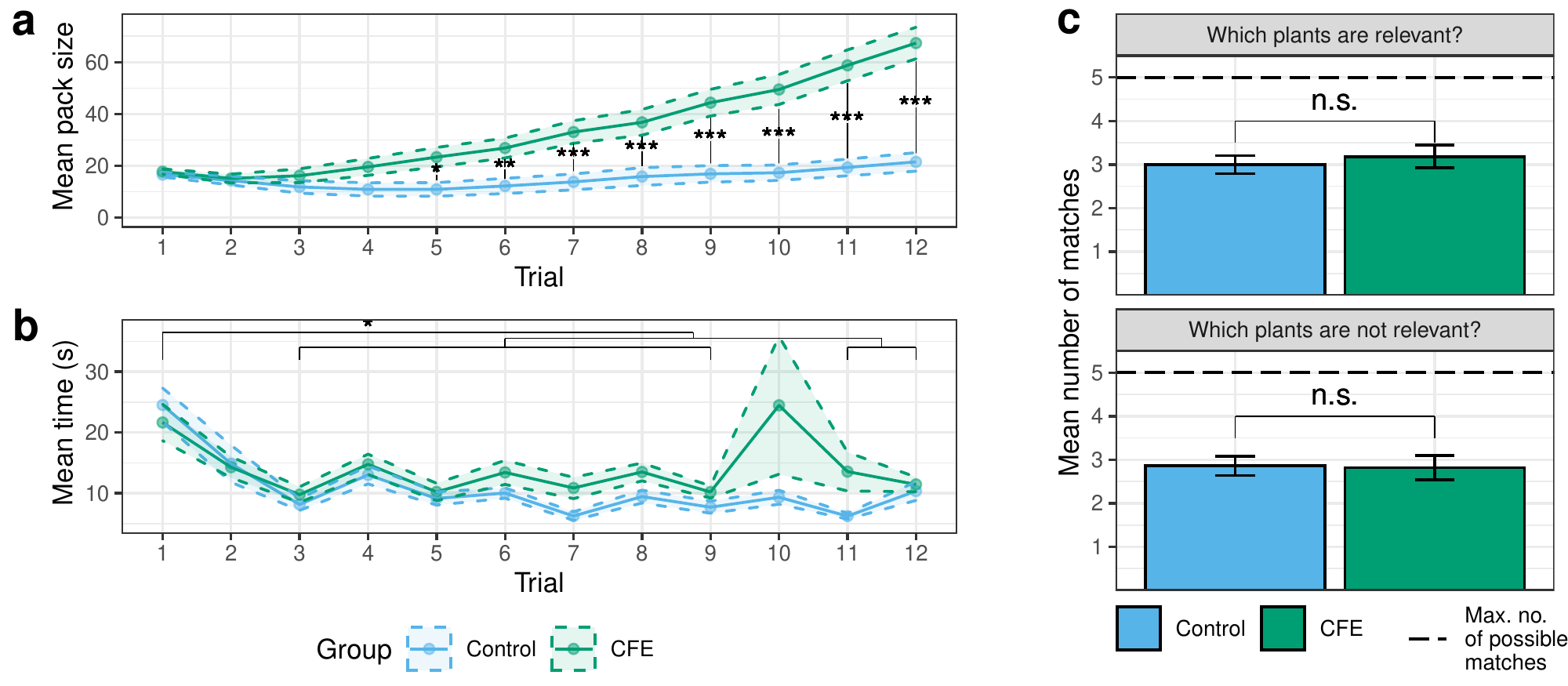}
   \caption{Experiment 2: Development of (\textbf{a}) mean pack size per group by trial, (\textbf{b}) mean decision time per group by trial, and (\textbf{c}) mean number of matches between user judgments and ground truth for survey items assessing relevant plants and irrelevant plants, respectively. Shaded areas in (\textbf{a}) and (\textbf{b}), and error bars in (\textbf{c}) denote the standard error of the mean. Asterisks denote statistical significance (\textit{p} < .05 (*), \textit{p} < .01 (**), and \textit{p} < .001 (***), respectively. n.s. = not statistically significant (i.e., \textit{p} > .05).}
   \label{fig:Exp2Hyp1}
 \end{figure*}

As in Experiment 1, average pack size per group increases significantly faster when CFEs are given (Figure \ref{fig:Exp2Hyp1}a; significant interaction of factors trial number and group; F(11,451) = 32.748, \textit{p} \textless .001, $\eta_{\text{p}}^{2}$ = 0.444). 
In contrast to Experiment 1, some users in the \textit{control} condition increases their pack size over the course of the experiment, in line with our expectation given the simpler data set.
Still, follow-up analyses reveal significant differences between groups from trial 5 onward (all t(67.8) $\geq$ 2.384, \textit{p} $\leq$ 0.020, \textit{d} $\geq$ 1.467).
Additionally, there is a significant main effect of trial number (\textit{F}(1,451) = 62.556, \textit{p} \textless .001, $\eta_{\text{p}}^{2}$ = 0.604), as well as a significant main effect of group (\textit{F}(1,41) = 16.909, \textit{p} \textless .001, $\eta_{\text{p}}^{2}$ = 0.292).

Similar to Experiment 1, participants in both groups showed a decrease in decision time over the curse of the study, evident after the very first trial (Figure \ref{fig:Exp2Hyp1}b).
A significant main effect of factor \textit{trial number} (\textit{F}(11,451) = 4.991, \textit{p} \textless .001, $\eta_{\text{p}}^{2}$ = 0.109) confirms this observation.
Corresponding post-hoc analyses show significant differences between trial 1 and all other trials (all \textit{t}(451) $\geq$ 3.432, \textit{p} $\leq$ .043, \textit{d} $\geq$ 1.740), except for trials 2 and 10.
Neither the main effect of factor \textit{group} (\textit{F}(1,41) = 2.758, \textit{p} = .104, $\eta_{\text{p}}^{2}$ = 0.063), nor the interaction between factors \textit{trial number} and \textit{group} (\textit{F}(11,451) = 1.439, \textit{p} = .152, $\eta_{\text{p}}^{2}$ = 0.034) reach significance.

Overall, these results support the initial findings from Experiment 1, emphasizing the beneficial role of providing \glspl{CFE} in the Alien Zoo for successful task completion.

\paragraph{Assessing user's explicit knowledge}

Unlike Experiment 1, there is no statistically meaningful difference between groups in terms of number of matches between user judgments of plant relevance for task success and the ground truth (\textit{control}: mean number of matches between user input and ground truth = 3.000 $\pm$ 0.207 \textit{SE}; \textit{CFE}: mean number of matches = 3.182 $\pm$ 0.260 \textit{SE}; \textit{U} = 255, \textit{p} = .554, \textit{r} = .090) 
as well as irrelevant plants (\textit{control}: mean number of matches between user input and ground truth = 2.857 $\pm$ 0.221 \textit{SE}; \textit{CFE}: mean number of matches = 2.819 $\pm$ 0.284 \textit{SE}; \textit{U} = 223.5, \textit{p} = .860, \textit{r} = .221), indicating greater success in building up explicit knowledge even without explanations, given the simpler data set. 

Thus, given the current data, the advantage of building better explicit knowledge when \glspl{CFE} are available seems to disappear.

\paragraph{Measures of Subjective Usability}

\begin{figure*}[t]
   \centering
   \includegraphics[width=\textwidth]{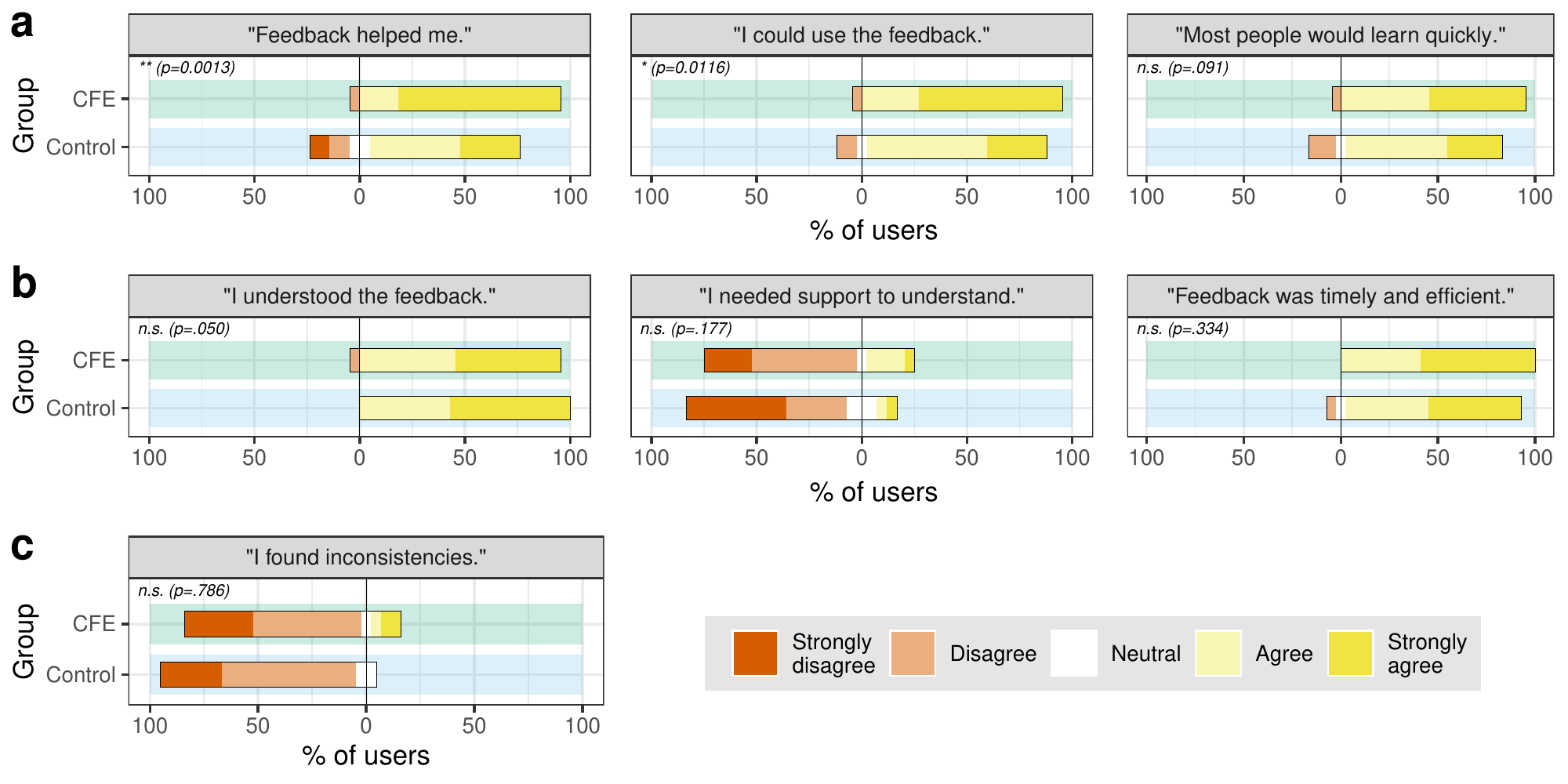}
   \caption{Experiment 2: Overview of user responses in post-game survey (adapted from ~\citep{holzinger_measuring_2020}) per group. (\textbf{a}) depicts user replies in survey items relevant for hypothesis 2, (\textbf{b}) depicts user replies in survey items relevant for hypothesis 3, and (\textbf{c}) depicts replies relevant for our last exploratory analysis. 
   Statistical information including the respective \textit{p}-value is given within each item's box (n.s. = not significant).}
   \label{fig:Exp2Survey}
 \end{figure*}

In stark contrast to Experiment 1, the majority of users from both groups in Experiment 2 shared a positive feeling that provided feedback was helpful and usable (\ref{fig:Exp2Survey}a).
The difference in response patterns still differs significantly between groups, in terms of subjective helpfulness (\textit{control} condition: \textit{M} = 3.714 $\pm$ 0.277 \textit{SE}; \textit{CFE} condition: \textit{M} = 4.682 $\pm$ 0.153 \textit{SE}; \textit{U} = 351, \textit{p} =  .001, \textit{r} = .489) and subjective usability (\textit{control} condition: \textit{M} = 4.048 $\pm$ 0.189 \textit{SE}; \textit{CFE} condition: \textit{M} = 4.591 $\pm$ 0.157 \textit{SE}; \textit{U} = 325, \textit{p} = .012, \textit{r} = .385). Extremely favorable user judgements from the \textit{CFE} group likely drive this effect, due to strong agreement by a large proportion of users from this cohort.

As in Experiment 1, there is no significant difference between groups for estimated usefulness of feedback for others (\textit{control} condition: \textit{M} = 3.952 $\pm$ 0.212 \textit{SE}; \textit{CFE} condition: \textit{M} = 4.409 $\pm$ 0.157 \textit{SE}; \textit{U} = 294.5, \textit{p} = .091, \textit{r} = .258).

\paragraph{Mode of Presenting Feedback and CFEs}
In accordance with Hypothesis 3, survey responses reflecting user's subjective understanding of feedback show that groups did not differ in terms of understanding the feedback as such (Figure \ref{fig:Exp2Survey}b; \textit{control} condition: \textit{M} = 4.571 $\pm$ 0.111 \textit{SE}; \textit{CFE} condition: \textit{M} = 4.409 $\pm$ 0.157 \textit{SE}; \textit{U} = 306, \textit{p} = .05, \textit{r} = .610).
Likewise, users in both groups indicate strongly that they do not wish for support to understand feedback provided (\textit{control} condition: \textit{M} = 1.905 $\pm$ 0.248 \textit{SE}; \textit{CFE} condition: \textit{M} = 2.318 $\pm$ 0.250 \textit{SE}; \textit{U} = 284, \textit{p} $\leq$ .177 \textit{r} = .206).
User judgments on timing and efficacy of presented feedback is consistently high across groups (\textit{control} condition: \textit{M} = 4.334$\pm$ 0.174 \textit{SE}; \textit{CFE} condition: \textit{M} = 4.591$\pm$0.107 \textit{SE}; \textit{U} = 266.5, \textit{p} = .334 \textit{r} = .147).

\paragraph{Identification of Inconsistencies}
Analysis of the final survey item reveals that users in both groups did not differ in finding inconsistencies in the feedback provided (\textit{control} condition: \textit{M} = 1.810 $\pm$ 0.131 \textit{SE}; \textit{CFE} condition: \textit{M} = 2.091 $\pm$ 0.254 \textit{SE}; \textit{U} = 241.5, \textit{p} = .786, \textit{r} = .041).

\subsection{Discussion}\label{sec:Disc}

In the empirical proof of concept study, we investigate the efficacy and feasibility of the Alien Zoo framework.
To this end, we examine the impact of providing \glspl{CFE} on user performance as compared to no explanations.
Based on objective behavioral variables and subjective self-reports, we assess understanding and usability of \gls{CFE}-style feedback.
Our results reveal the potential of the Alien Zoo framework to study the usability of \glspl{CFE} approaches.

Most notably, merely providing a summary of past choices does not necessarily enable users to gain insight into the system. 
This becomes especially clear considering the poor task performance of \textit{control} participants in the more complex Experiment 1.
Given the comparatively complex interdependence of three features in the underlying data, none of the \textit{control} participants manage to increase their pack size in the course of the experiment.

Participants receiving \glspl{CFE} for their choices, however, are able to manipulate the system more efficiently. 
While both experiments vary in terms of the complexity of the underlying data used for model building, the observation of \glspl{CFE} participants outperforming their peers in the \textit{control} group, is consistent.
Interestingly, the \textit{control} group in Experiment 2 did indeed manage to improve their pack size to some extent, but providing explanations puts users at a definite advantage. In fact, 100\% of all users in the experimental condition in Experiment 2 correctly determine that plant 2 is a relevant feature. 
This observation not only supports the claim that \glspl{CFE} are a very intuitive and meaningful way of explaining in \gls{XAI}~\citep{wachter2017counterfactual}, but clearly demonstrates their effectiveness in the current setting.

Intriguingly, our results diverge from those of empirical \gls{XAI} studies that find no beneficial effect of providing \glspl{CFE} on user's task performance~\citep{lim_why_2009, van_der_waa_evaluating_2021}.
For instance, ~\citeauthor{lim_why_2009} review various explanation approaches in the domain of context-aware systems~\citep{lim_why_2009}.
Their evidence suggests that users receiving counterfactual style \textit{what-if}-explanations have no advantage over control users when manipulating abstract features (labelled \textit{A}, \textit{B} and \textit{C}) to explore their influence on abstract predictions (labelled \textit{a} or \textit{b}).

In an attempt to explain this stark contrast to our results, we may turn to the details of both experimental tasks. 
First, the Alien Zoo revolves around an engaging setting (\ie, feeding aliens to make the pack grow), as opposed to the non-specific nature of the system in~\citeauthor{lim_why_2009}.
Second, we offer users different rounds of action and feedback in alternating learning and testing steps, making the Alien Zoo truly interactive.
In contrast, users in~\citeauthor{lim_why_2009} undergo an initial evaluation section displaying explanation after explanation, followed by a separate test phase.
Learners obtaining deeper understanding through hands-on activities rather than passive studying is well established in educational science~\citep{chi_icap_2014}, potentially explaining discrepancies in terms of observed user behavior.
Thus, we suggest that future \gls{XAI} usability studies should put a strong focus on goal-directed and interactive tasks to be maximally effective.

All users across conditions and experiments significantly decrease the time needed to reach a decision. This decrease becomes apparent already after the very first trial, most likely reflecting how participants initially take some moments to familiarize themselves with the interface.
Another slight increase is observable for trial 4, right after the first in-game attention question appeared. 
We assume that users took this trial to focus their concentration again after this surprising disruption. 
From there on, decision times consistently level out for both groups.
Thus, despite the performance benefit, we have no evidence that providing \glspl{CFE} leads to more automatic, and thus faster, decision-making.

In the more complex Experiment 1, users in the experimental group can more correctly state which plants are relevant for the task, compared to users in the control group. 
Interestingly, in the simpler Experiment 2, this significant difference vanishes.
This might reflect the greater success of control users to see through the system in this simpler setting, even without explanations.
However, this should not be taken as evidence that users across groups build up mental models of the underlying system that are indeed comparable, considering the considerable difference in task performance. 
In fact, one caveat of the current analysis may be insufficient sensitivity of the measure of matches between user input and ground truth, possibly diluting noteworthy effects.
For instance, 100\% of all users in the experimental condition, but only 57\% of all control participants, could determine that plant 2 is a relevant feature in Experiment 2 (see Supplementary Material \ref{app:suppRelevanceJudgm}). The current measure does not capture this detail, calling for careful interpretation of the corresponding null-effect.

On top of the objective measures quantifying system understanding, we assess various subjective measures to tap into perceived usability.
Across both experiments, the experimental groups judged their \gls{CFE}-style feedback as being more helpful and usable compared to the control group (Figures \ref{fig:Exp1Survey}a and \ref{fig:Exp1Survey}a, respectively). 
Thus, providing \glspl{CFE} does not just improve user's performance, but also their subjective usability of the system.

Surprisingly, despite variable responses in terms of helpfulness and usability of presented feedback for oneself, the estimated usefulness for others is not different across groups. 
In fact, a larger proportion of control users in Experiment 1 reported favorably on that item, even though they found feedback of little help and limited usefulness.
This astonishing result is difficult to interpret without access to more detailed qualitative data from those participants. 
Maybe these users are demotivated by their poor turnout, feeling that they perform exceptionally bad compared to the average person.

Participant responses to items in place to assess potential confounding factors reveal an interesting pattern that merits closer inspection (Figures \ref{fig:Exp2Survey}b and \ref{fig:Exp2Survey}b).
In Experiment 1, a considerable proportion of both groups responds positively about understanding the feedback. 
However, the \textit{control} group leans significantly more towards agreement.
This might reflect higher cognitive load the \gls{CFE} group, as they receive a more crowded, information-heavy screen. 
While in line with findings suggesting that counterfactual style questions impose a larger cognitive load on participants~\citep{lage2019human}, this interpretation is unlikely as this effect vanishes in the simpler Experiment 2.
This fact rather suggests that the increased task difficulty drives this effect.
Response patterns on the other two control items are more consistent.
Users across groups and across experiments state that they need little support to understand the feedback provided.
Similarly, an overwhelming majority of all users across all groups indicate that feedback was timely and efficient, backing the efficacy of the Alien Zoo framework despite its relatively complex game-like setup.

Survey items depicted in Figures \ref{fig:Exp2Survey}b and \ref{fig:Exp2Survey}b are set in place to assess potential confounding factors, possibly impacting the efficiency of the Alien Zoo framework.
We assumed that across experiments, consistent group difference with respect to these items would inform us about potential design flaws.
While one group difference emerges, however, it is not consistent across experiments. 
This clearly indicates that the respective item (\textit{``I understood the feedback.''}) not just evaluates general understanding, but also reflects the underlying task difficulty.
A possible explanation for this may be that there is still room for improvement for a clean identification of confounds.
In lack of a standard inventory for assessing subjective usability in \gls{XAI} user studies, we rely on an adapted version of the System Causability Scale \cite{holzinger_measuring_2020}.
While a very good starting point, it may be a worthwhile endeavor to perfect this measure in future user validations.

Finally, our exploratory analysis reveals that groups in both experiments do not differ in finding inconsistencies in the feedback provided.
This acts as a further quality measure for the \gls{CFE} approach, trusted to generate feasible and sound explanations. 
While this verdict is virtually unanimous across users in the simpler Experiment 2, some users in both groups in Experiment 1 indicate that they did indeed determine inconsistencies.
While a minority, this observation merits a comment.
We cannot exclude that some users in the \gls{CFE} group indeed receive feedback in different runs that, when taken together, does not perfectly align. 
It is important to keep in mind that \glspl{CFE} are local explanations, highlighting what would lead to better results in a particular instance.
Variability, especially in terms of the irrelevant features, may indeed exist.
To uncover whether such effects cause fundamental problems, we intentionally moved away from the perfect, hand-crafted explanations assessed classical \textit{Wizard of Oz} designs more prominently used in the community~\citep{lage2019human,narayanan2018humans, sokol2020one,van_der_waa_evaluating_2021}, and used predictions from real ML-models.
However, the observation that a small proportion of users in the control group indicate that they found inconsistencies, is much more puzzling.
These users merely see a correct summary of their past choices as feedback, and thus inconsistencies are impossible.
Given that this survey item was the very last, it may be a sign of participants' loss of attention or fatigue.
Identifying the actual underlying reasons requires collecting quantitative data, \eg, via in depth user interviews. 
These measurements require moving away from the accessible web-based format, and perform complementary evaluations in an in-person, lab-based setting.

\subsubsection{Limitations}

Several limitations to this proof of concept study deserve discussion.
While we clearly demonstrate the benefit of providing \glspl{CFE} as feedback in an iterative learning design targeting an abstract domain for novice users, generalization of this observation to other tasks, domains and target groups is extremely limited\citep{doshi-velez_towards_2017,sokol2020explainability}.

A cautionary note regards the efficacy of \glspl{CFE} for human users more generally. 
\glspl{CFE} are local explanations, focusing on how to undo one past prediction. 
Thus, it is very unlikely that users are able to form an accurate mental model of the entire underlying system solely based on a sparse set of these specific explanations. 
This is a short-coming, given that completeness is an important prerequisite for this process~\citep{kulesza2013too}.
Thus, it remains an avenue for future research to show situations that severely impact usability of \glspl{CFE}, as they are unable to provide a complete picture.

Another point to keep in mind is the potential problem of users falling victim to confirmation bias after receiving the first round of \gls{CFE} feedback~\citep{wang2019designing}. 
In essence, we cannot rule out that some users generate a faulty initial hypothesis, and subsequently look for confirming evidence for that faulty initial hypothesis only. 
This may have greater impact on the control group, given that they have very little evidence to go by choosing the best plant combination. 
Still, it also needs to be acknowledged as a possible issue for the \gls{CFE} participants.
Consequently, such a strategy would hamper learning profoundly, and we cannot rule out that some lower performing users indeed follow it.
While exploring the impact of confirmation bias for \glspl{CFE} in \gls{XAI} is outside the scope of this work, the issue deserves more careful attention in future work.

Finally, we do not investigate whether providing \glspl{CFE} did also improve user's trust in the system.
Trust is an important factor in XAI, and prominently studied in various designs~\citep{davis2020measure,lim_why_2009, ribeiro2016should}.
The current work, however, exclusively focuses on the aspect of usability.
Extending the current set up to include evaluation of trust can be easily realized, for instance by extending the survey by corresponding items.

Finally, a further insight gained from this study is the critical impact of task difficulty on user performance and judgements.
While not directly at the center of the current work, we shed a first light on these effects by observing differences between Experiment 1 and 2.
Future research should look into the effects of data complexity on usability of \glspl{CFE}.

\subsubsection{Conclusions}

The main contributions of the empirical proof of concept study are two-fold.
First, we provide long-awaited empirical evidence for the claim that \glspl{CFE} are indeed more beneficial for users than providing no explanations, at least in abstract setting, when tasked to gain new knowledge.
Importantly, this advantage becomes apparent both in terms of objective performance measures and subjective user judgements.
Second, we demonstrate the basic efficacy of the Alien Zoo framework for studying the usability of \glspl{CFE} in \gls{XAI}.

\section{Future Perspectives and Conclusions}\label{sec:DiscConcl}

The current paper introduces the Alien Zoo framework, developed to assess the usability of \glspl{CFE} in \gls{XAI}.
In a proof of concept study, we demonstrate its efficacy by examining the added benefit of providing \glspl{CFE} over no explanations using an iterative learning task in the abstract Alien Zoo setting.

User evaluations of \gls{XAI} approaches are still in their infancy, leaving abundant room for studying various aspects.
We believe that the Alien Zoo enables researchers to investigate a wide variety of different questions.
For instance, in a separate study, we use the Alien Zoo to investigate potential advantages of \glspl{CFE} restricted to plausible regions of the data space compared to classical \glspl{CFE} remaining as close to the original input as possible~\citep{kuhl2022keep}. Surprisingly, this investigation reveals that novice users in the current task do not benefit from an additional plausibility constraint.

Another issue for future research may be to examine usability of different types of \glspl{CFE}. Importantly, \glspl{CFE} may vary in terms of framing the respective result. Upward counterfactuals highlight how the current situation would be improved, while downward counterfactuals emphasize changes leading to a less desirable outcome~\citep{epstude_functional_2008}. The impact of such a framing in \gls{XAI} is yet to be shown.

Moreover, further research should be done to uncover potential differences in usability for \glspl{CFE} generated for different models. While the way \glspl{CFE} are presented in the Alien Zoo is always the same, the underlying models may be fundamentally different. Thus, if human users pick up on model differences solely based on their respective explanations, it may have critical implications for their usability. A particularly intriguing question to be addressed is whether users are able to identify a model that is objectively worse.

As a final suggestion of this by no means exhaustive list, we propose studying potentially negative effects of \glspl{CFE}: In the field of \glspl{XAI}, it is universally assumed that \glspl{CFE} are intuitive and human-friendly. Thus, it will be extremely informative to investigate and identify cases where these types of explanation do more harm than good, \eg, when users come to trust \gls{ML} models even if they are biased and unfair.

It is natural for people to interact with each other by explaining their behaviors to one another.
The key to building a stable mental model for prediction and control of the world is to explain in a way that is understandable and usable~\citep{heider_psychology_1958}.
However, in the absence of a universally applicable definition of what constitutes a good explanation, a lack of user-based evaluations affects the assessment of automatically generated \glspl{CFE} for \gls{ML}.

The lack of user-based research does not only bear upon assessments of \glspl{CFE} as such, but also limits the overall evaluation of different conceptualizations for this kind of explanations.
Consequently, with the Alien Zoo framework, we offer a flexible, easily adaptable design, applicable for various purposes and research questions. This approach in its implementation may be freely used by researchers and practitioners to further advance the field of \gls{XAI}.

\AtNextBibliography{\raggedright\small}
\printbibliography


\newpage
\onecolumn

\appendixpageoff
\appendixtitleoff
\renewcommand{\appendixtocname}{Supplementary material}

\begin{appendices}
    \section*{Supplementary material}
    \section{Full List of Survey Items}
\label{app:suppSurvey}

\begin{table}[htb]
\caption*{* marks the catch item in place to evaluate if users are still paying attention.}
\label{tab:suppSurvey}
\resizebox{\textwidth}{!}{%
\begin{tabular}{@{}llll@{}}
\toprule
Item No. &
  Control group &
  CFE group &
  Response options \\ \midrule
1 &
  \multicolumn{2}{l}{\begin{tabular}[c]{@{}l@{}}What do you think: Which plants were relevant to increase\\ the number of Shubs in your pack?\\ Please select ALL that you think were relevant.\end{tabular}} &
  \multirow{2}{*}{\begin{tabular}[c]{@{}l@{}}5 checkboxes, together with icons\\ of the available plants \\ + option "I do not know."\end{tabular}} \\
2 &
  \multicolumn{2}{l}{\begin{tabular}[c]{@{}l@{}}What do you think: Which plants were not relevant to increase \\ the number of Shubs in your pack?\\ Please select ALL that you think were not relevant.\end{tabular}} &
   \\ \midrule
3 &
  \begin{tabular}[c]{@{}l@{}}I understood the overview\\ of my past choices.\end{tabular} &
  \begin{tabular}[c]{@{}l@{}}I understood the feedback\\ on what choice would have\\ led to a better result.\end{tabular} &
  \multirow{8}{*}{\begin{tabular}[c]{@{}l@{}}5 point Likert-scale,\\ checkboxes with options: \\ Strongly disagree - disagree - neutral \\ - agree - strongly agree \\ + option "I prefer not to answer."\end{tabular}} \\
4 &
  \begin{tabular}[c]{@{}l@{}}I needed support to understand \\ the overview of my past choices.\end{tabular} &
  \begin{tabular}[c]{@{}l@{}}I needed support to understand\\ the feedback on what choice\\ would have led to a better result.\end{tabular} &
   \\
5 &
  \begin{tabular}[c]{@{}l@{}}I found that the overview of my\\ past choices helped me to increase\\ the number of Shubs.\end{tabular} &
  \begin{tabular}[c]{@{}l@{}}I found that the feedback on\\ what choice would have led\\ to a better result helped me\\ to increase the number of Shubs.\end{tabular} &
   \\
6 &
  \begin{tabular}[c]{@{}l@{}}I was able to use the overview\\ of my past choices to increase\\ the number of Shubs.\end{tabular} &
  \begin{tabular}[c]{@{}l@{}}I was able to use the feedback\\ on what choice would have led\\ to a better result to increase\\ the number of Shubs.\end{tabular} &
   \\
7* &
  \multicolumn{2}{l}{\begin{tabular}[c]{@{}l@{}}Are you still paying attention? If so, please select 'I prefer not to answer'\\ for this question.\end{tabular}} &
   \\
8 &
  \begin{tabular}[c]{@{}l@{}}I found inconsistencies in the\\ overview of my past choices.\end{tabular} &
  \begin{tabular}[c]{@{}l@{}}I found inconsistencies in the\\ feedback on what choice\\ would have led to a better result.\end{tabular} &
   \\
9 &
  \begin{tabular}[c]{@{}l@{}}I think most people would learn\\ to work with the overview of \\ their past choices very quickly.\end{tabular} &
  \begin{tabular}[c]{@{}l@{}}I think most people would learn\\ to work with the feedback\\ on what choice would have\\ led to a better result very quickly.\end{tabular} &
   \\
10 &
  \begin{tabular}[c]{@{}l@{}}I received the overview of my\\ past choices in a timely and\\ efficient manner.\end{tabular} &
  \begin{tabular}[c]{@{}l@{}}I received the feedback on what\\ choice would have led to a better\\ result in a timely and efficient manner.\end{tabular} &
   \\ \midrule
Age &
  \multicolumn{2}{l}{Please indicate your age.} &
  \begin{tabular}[c]{@{}l@{}}Checkboxes with options: \\ 18-24y, 25-34y, 35-44y, \\ 45-54y, 55-64y, 65y and over\end{tabular} \\ \midrule
Gender &
  \multicolumn{2}{l}{Which term most accurately describes your gender?} &
  \begin{tabular}[c]{@{}l@{}}Checkboxes with options: \\ Female, Male, \\ Transgender female, Transgender male, \\ Non-binary / gender non-conforming, \\ Not listed, I prefer not to answer\end{tabular} \\ \bottomrule
\end{tabular}}
\end{table}
    \newpage
    \section{Exemplary User Journey Through the First Block of the Game Phase}
\label{app:suppSceneFlow}

\begin{figure}[htb]
   \centering
   \includegraphics[width=\textwidth]{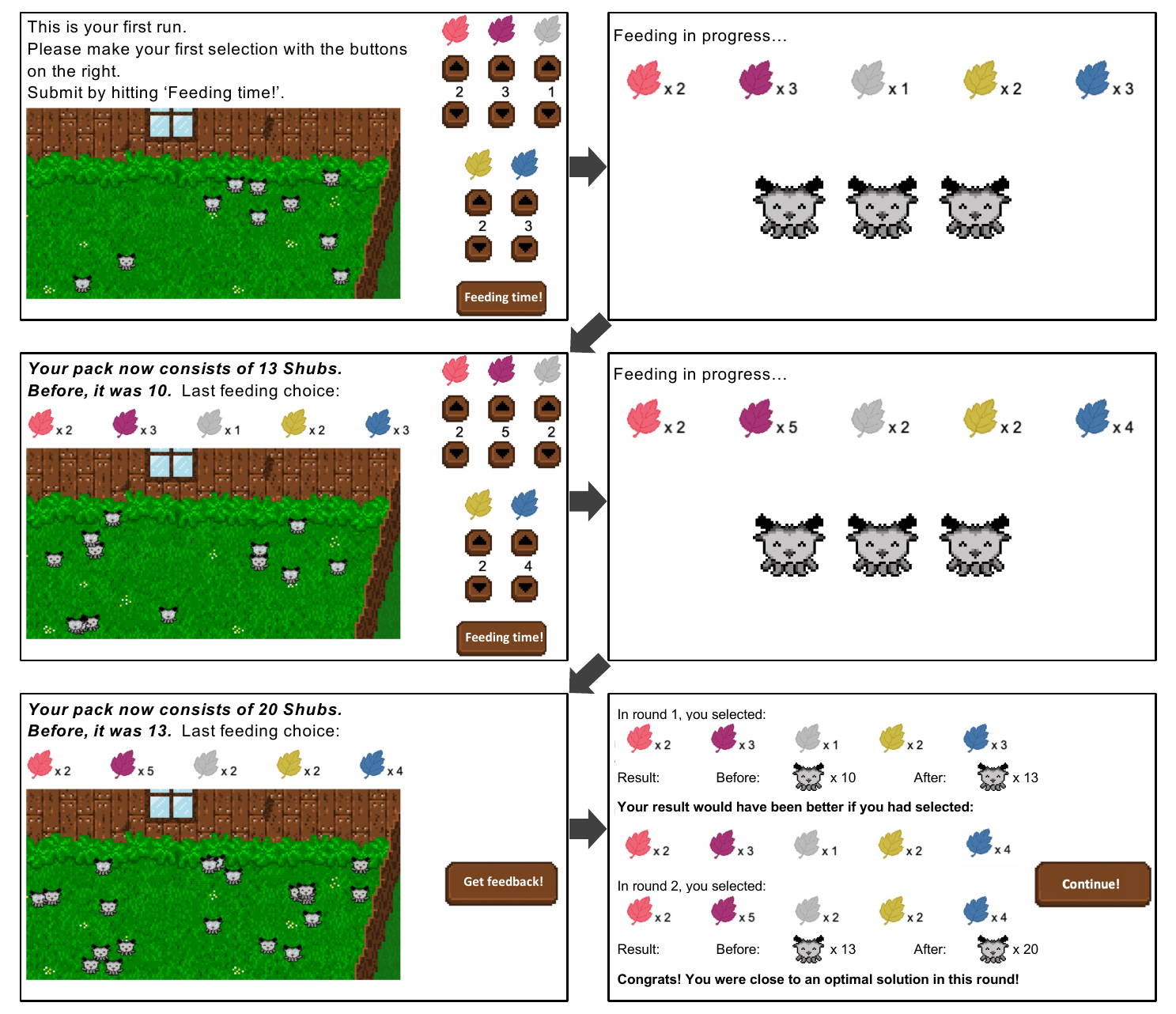}
   \caption*{Exemplary user journey through the first block of the Alien Zoo game. Bold arrows indicate temporal succession of respective scenes. The figure highlights the iterative nature of the game with repeated user input and end-of-block presentation of \glspl{CFE} (experimental group), or overview of past choices (control group). Note that plant counters are set to 0 at the beginning of each padlock scene. The figure displays the state after the exemplary user inserted their current choice. For this manuscript, font size in images of scenes was increased to improve visibility.}
   \label{fig:suppSceneFlow}
 \end{figure}

    \newpage
    \section{Detailed Feature Relevance Judgements}
\label{app:suppRelevanceJudgm}

\begin{figure}[htb]
   \centering
   \label{fig:suppRelevanceJudgmExp1}
   \includegraphics[width=0.75\textwidth]{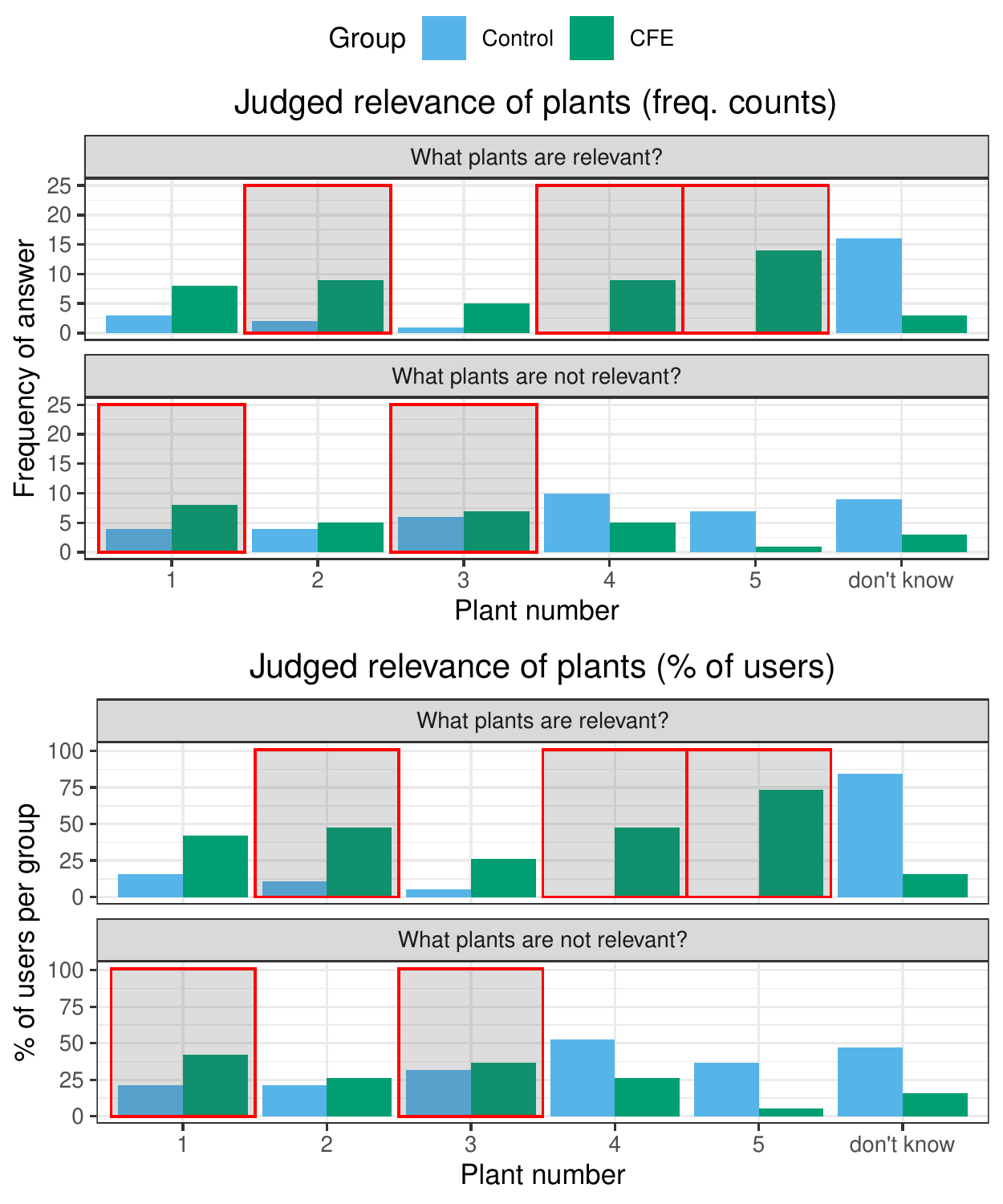}
   \caption*{Experiment 1: Detailed relevance judgements per plant, as frequency of users (top), and percentage of users (bottom), respectively. Red boxes indicate relevant and non-relevant features.}
 \end{figure}
 
 \begin{figure}[htb]
   \centering
   \label{fig:suppRelevanceJudgmExp2}
   \includegraphics[width=0.75\textwidth]{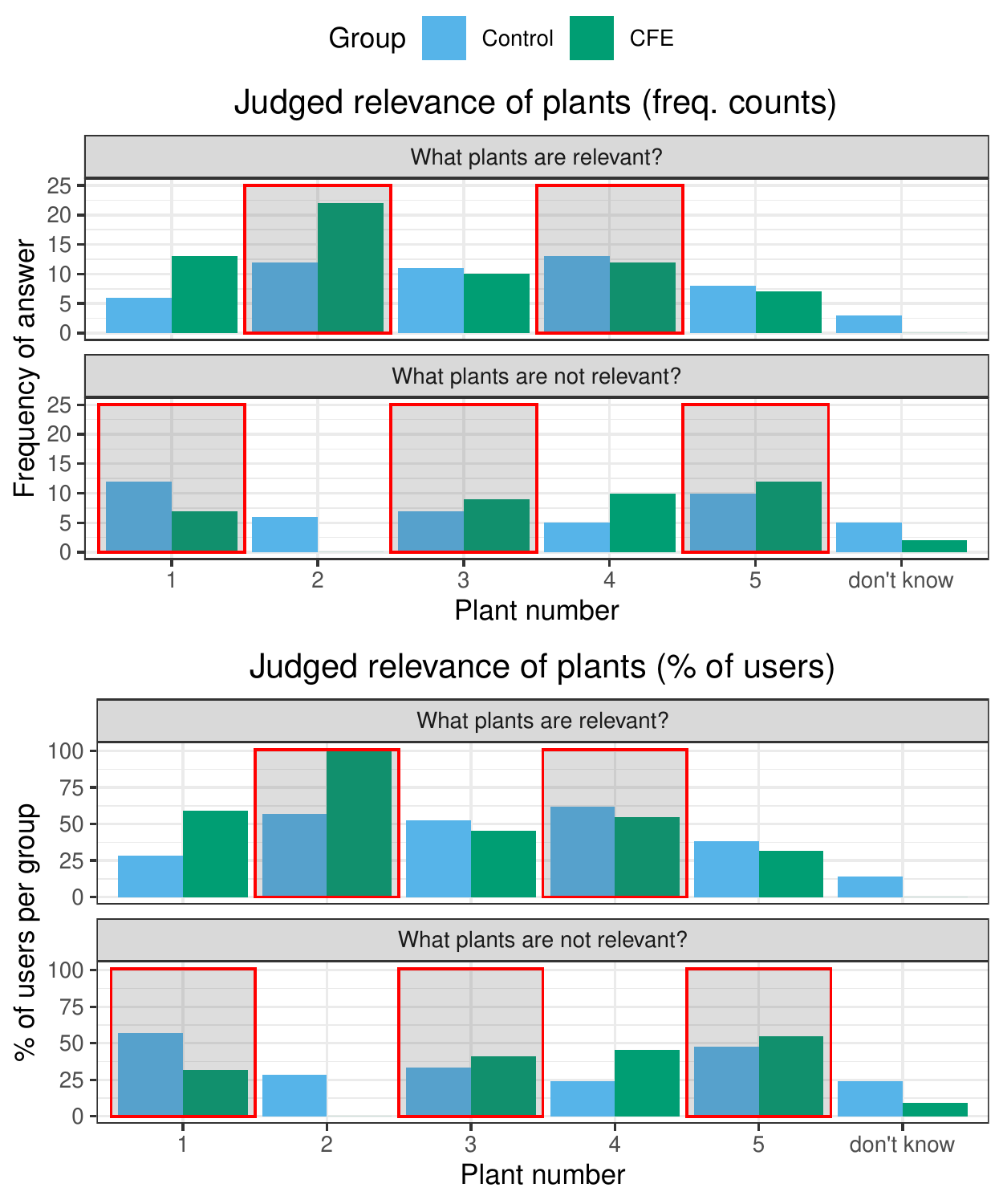}
   \caption*{Experiment 2: Detailed relevance judgements per plant, as frequency of users (top), and percentage of users (bottom), respectively. Red boxes indicate relevant and non-relevant features.}
 \end{figure}
 
\end{appendices}

\end{document}